\documentclass[conference]{IEEEtran}
\IEEEoverridecommandlockouts
\usepackage{cite}
\usepackage{caption}
\usepackage{subcaption}
\usepackage{amsmath,amssymb,amsfonts}
\usepackage{algorithmic}
\usepackage{graphicx}
\usepackage{multirow}
\usepackage{multicol}
\usepackage{textcomp}
\usepackage{xcolor}
\def\BibTeX{{\rm B\kern-.05em{\sc i\kern-.025em b}\kern-.08em
    T\kern-.1667em\lower.7ex\hbox{E}\kern-.125emX}}
\begin{document}

\title{Assessing the Impact of Low Resolution Control Electronics on Quantum Neural Network Performance}

\author{\IEEEauthorblockN{Rupayan Bhattacharjee, Rohit Sarma Sarkar, Sergi Abadal, Carmen G. Almud\'ever$^{1}$, Eduard Alarc\'on}
\IEEEauthorblockA{\textit{Nanonetworking Center in Catalonia (N3Cat), Universitat Polit\`{e}cnica de Catalunya, Spain}
\\ $^1$\textit{Universitat Polit\`ecnica de Valencia, Spain}
\\
Email: rupayan.bhattacharjee@upc.edu}
}

\maketitle

\begin{abstract}
Scaling quantum computers requires tight integration of cryogenic control electronics with quantum processors, where Digital-to-Analog Converters (DACs) face severe power and area constraints. We investigate quantum neural network (QNN) training and inference under finite DAC resolution constraints, evaluating two QNN architectures across four diverse datasets (MNIST, Fashion-MNIST, Iris, Breast Cancer). Pre-trained QNNs achieve accuracy nearly indistinguishable from infinite-precision baselines when deployed on quantum systems with 6-bit DAC control electronics, exhibiting characteristic elbow curves with diminishing returns beyond 3-5 bits depending on the dataset. However, training QNNs directly under quantization constraints reveals gradient deadlock below 12-bit resolution, where parameter updates fall below quantization step sizes, preventing training entirely. We introduce temperature-controlled stochastic quantization that overcomes this limitation through probabilistic parameter updates, enabling successful training at 4-10 bit resolutions. Remarkably, stochastic quantization not only matches but frequently exceeds infinite-precision baseline performance across both architectures and all datasets. Our findings demonstrate that low-resolution control electronics (4-10 bits) need not compromise QML performance while enabling substantial power and area reduction in cryogenic control systems, presenting significant implications for practical quantum hardware scaling and hardware-software co-design of QML systems.
\end{abstract}


\section{Introduction}

Quantum Machine Learning (QML) leverages quantum mechanical systems to enhance machine learning tasks \cite{biamonte2017quantum, schuld2015introduction}, offering potential speedups over classical approaches for specific problems \cite{havlivcek2019supervised, liu2021rigorous, glick2024covariant, huang2021power, huang2022quantum, huang2025generative}. QML has demonstrated promise across diverse domains including image processing \cite{senokosov2024quantum, chen2024novel, SUN2025130226}, finance \cite{mironowicz2024applications, mancilla2022preprocessing}, and drug discovery \cite{doi:10.1021/acs.jcim.1c00166, doi:10.1021/acs.chemrev.4c00678}. Quantum Neural Networks (QNNs), particularly variational quantum circuits, represent a leading paradigm for implementing QML on near-term Noisy Intermediate-Scale Quantum (NISQ) devices \cite{preskill2018quantum, cerezo2021variational}.

Scaling quantum computers for practical QML applications necessitates tight integration of cryogenic CMOS control electronics with quantum processors \cite{PhysRevApplied.12.044054, gonzalez2021scaling}. These control systems face severe constraints: limited power budgets and restricted chip area \cite{8036394, 10.1145/3061639.3072948}. A critical bottleneck lies in the Digital-to-Analog Converters (DACs/D2As) that generate control pulses for quantum gate operations. Higher DAC precision (increased bit depth) demands greater power consumption and silicon area \cite{8036394}, creating fundamental trade-offs in hardware design.

Prior work has explored related but distinct aspects of this challenge. Probabilistic gate synthesis methods \cite{koczor2024probabilistic, koczor2024sparse} achieve exact gate implementation on low-resolution hardware through post-processing techniques, but incur substantial computational overhead. A QNN compression approach proposed by \cite{hu2022quantum} reduces circuit complexity through pruning and quantization, focusing on minimizing transpiled circuit depth and gate count rather than addressing QML training and inference with control electronics limitations. Neither line of work examines how finite DAC resolution fundamentally constrains the training process itself, specifically, how DAC quantization affects parameter updates during gradient-based optimization and how the training/inference capability of QNNs is affected by DAC resolution. The interplay between control electronics precision and quantum algorithm performance remains an open question.

We systematically address this by investigating QNN performance under realistic DAC resolution constraints, evaluating two architectures across four diverse classification tasks. For inference, pre-trained QNNs trained with infinite precision are tested with low-resolution DACs. Additionally, we investigate the training of QNNs with quantization limitations, revealing a critical bottleneck, a \textit{gradient deadlock} phenomenon that inhibits effective parameter updates in the QNN. Furthermore, we introduce temperature-controlled stochastic quantization to overcome gradient deadlock during training, explicitly examining QML with control electronics constraints.

The main contributions of this work are:
\begin{itemize}
    \item  Evaluation of inference accuracy of pre-trained QNNs on systems with finite-resolution DACs.
    \item Temperature-controlled stochastic parameter updates that enable QNN training with low-resolution DACs, overcoming gradient deadlock.
    \item Demonstration that low-resolution systems can match or exceed infinite-precision QNN performance, enabling practical hardware-software co-design of QML systems.
\end{itemize}
Our results challenge the assumption that quantum control requires maximum precision, enabling practical QML deployment on resource-constrained quantum systems and bridging the gap between algorithmic requirements and hardware capabilities for near-term quantum advantage.

\begin{figure*}[t]
    \centering
    \includegraphics[width=\textwidth]{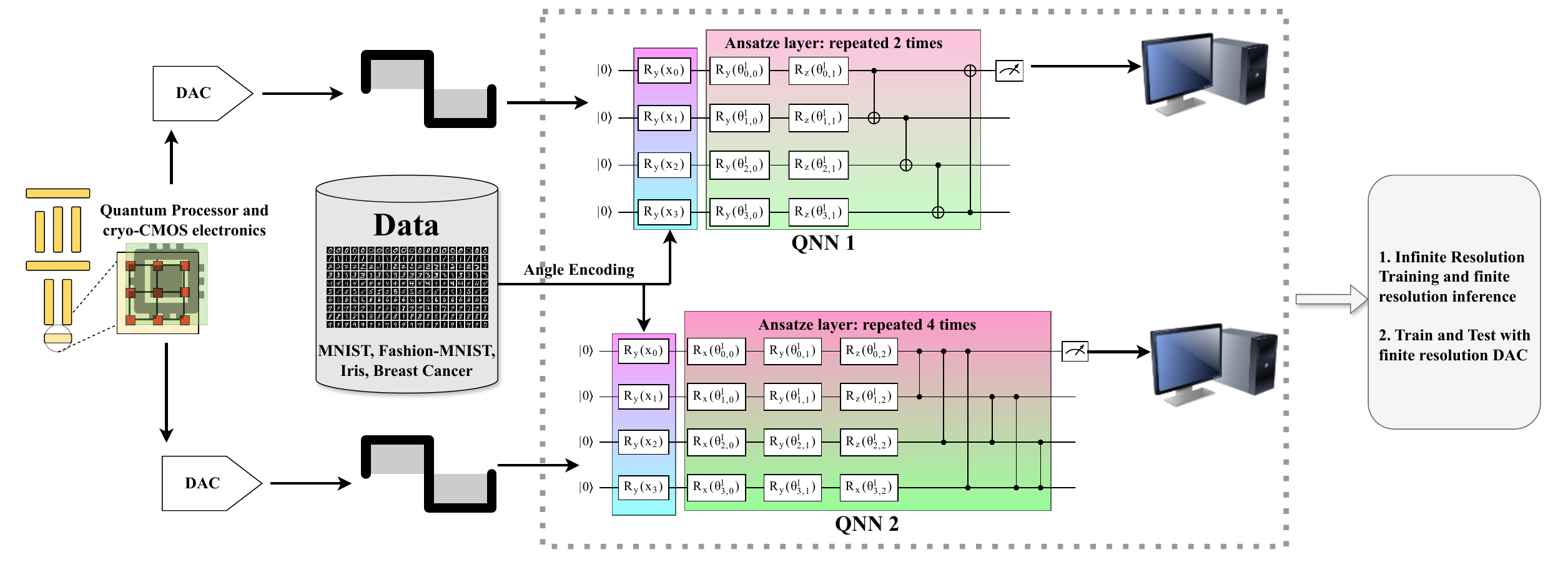}
    \vspace{-0.5cm}
    \caption{Methodology workflow.}
    \label{fig:methodology_figure}
\end{figure*}

\section{Background}

This work lies at the intersection of QML and cryogenic quantum control electronics. In this section, we provide essential background to make our contributions accessible to the broader computing systems community.

\subsection{Quantum Neural Networks}

QNNs are parameterized quantum circuits that process information encoded in qubits \cite{cerezo2021variational}. A QNN implements a quantum circuit $U(\vec{x})$ acting on the initial state $|0\rangle^{\otimes N}$ to encode classical data $\vec{x}$ into a quantum state $|\Psi(\vec{x})\rangle$, inducing an implicit feature map. The feature map typically consists of Pauli gates parametrized by the normalized values of the data features. This is followed by the layerwise application of a parametrized quantum circuit (ansatze) $V(\vec{\theta})$, where each ansatze layer is constructed via the application of single-qubit Pauli gates, e.g. $R_{j\in \{x,y,z\}}(\theta)=exp(-i\frac{\theta}{2}\sigma_{j\in \{x,y,z\}})$ (where $\sigma_j$s are the Pauli matrices), and two-qubit entangling gates, e.g. CNOT, CZ, etc. The ansatze parameters $\vec{\theta}$ serve as the trainable weights of the QNN. The variational parameters are varied to optimize expectation value of an observable $\hat{O}$ which serves as the loss function $\mathcal{L}(\vec{x},\vec{\theta})=\langle \Psi(\vec{x},\vec{\theta})|\hat{O}|\Psi(\vec{x},\vec{\theta})\rangle=\langle0|^{\otimes N}U^{\dagger}(\vec{x})V^{\dagger}(\vec{\theta})\hat{O}V(\vec{\theta})U(\vec{x})|0\rangle^{\otimes N}$. Training is done via gradient descent (i.e. $\theta \leftarrow\theta - \eta \nabla_{\theta}\mathcal{L}$). On classical simulators, gradients are computed via automatic differentiation. On real quantum hardware, the parameter-shift rule is used to evaluate gradients, i.e. $ \nabla_\theta \mathcal{L} = \frac{1}{2}[\mathcal{L}(\theta_i + \pi/2) - \mathcal{L}(\theta_i - \pi/2)]$ \cite{schuld2019evaluating}, which implements the QNN circuit at two offset parameter values. The distribution of two-qubit gates in the circuit, called entangling strategy, is known to determine the expressivity and entangling capacity of QNNs \cite{sim2019expressibility}.

\subsection{Quantum Control Electronics}

Quantum processors require precise classical control systems to manipulate qubits. Across quantum hardware platforms, DACs generate analog voltage waveforms that implement quantum gates. Each rotation gate $R_j(\theta)$ requires setting a voltage proportional to angle $\theta$. For an $N$-bit DAC, angles are constrained to $2^N$ discrete levels with quantization step size $\Delta = 2\pi/(2^N - 1)$. Higher DAC resolution provides finer angle control but incurs significant increases in power consumption and chip area. For quantum systems with hundreds of qubits requiring multiple DAC channels per qubit, aggregate DAC power and area become critical scalability bottlenecks \cite{8036394}. Additionally, qubits operate at deep cryogenic temperatures where cooling power is extremely limited \cite{10287646}. Excessive power consumption by cryo-CMOS circuits, generates heat that introduces noise and degrades qubit states, rendering them useless for reliable computation. This tradeoff between gate precision and power consumption, given the limited power budget, motivates a key question: \textit{What is the minimum DAC resolution required for practical QNN operation?}

\section{Methodology}

We systematically investigate QNN performance under DAC quantization constraints through two experimental paradigms: (1) \textit{inference with post-training quantization}, where pre-trained QNNs (trained with infinite precision) are deployed on systems with finite-resolution DACs, and (2) \textit{training with quantization}, where QNNs are trained from scratch with quantization constraints enforced throughout optimization. This dual approach enables us to separately assess inference robustness and training feasibility under hardware constraints. The complete methodology workflow is illustrated in Figure~\ref{fig:methodology_figure}.

\subsection{Datasets}

We evaluate two 4-qubit QNNs for binary classification across four diverse datasets: handwritten digit recognition (MNIST \cite{deng2012mnist}, digits 0 vs. 1), clothing categorization (Fashion-MNIST \cite{DBLP:journals/corr/abs-1708-07747}, T-shirt/top vs. trouser), botanical classification (Iris \cite{Fisher1936THEUO}, Setosa vs. Versicolor), and medical diagnosis (Wisconsin Breast Cancer \cite{Street1993NuclearFE}, malignant vs. benign). For MNIST and Fashion-MNIST, we use 400 samples with a 70\%-30\% train-test split, reducing the dataset size for efficient training of small-scale QNNs (without risking inflating performance metrics \cite{phalak2023shot}), consistent with standard practice in QML literature \cite{havlivcek2019supervised, phalak2023shot, perez2020data, bowles2024better, acedo2025pulsed}. The Iris dataset contains 100 samples (50 per class) with the same split ratio. The breast cancer dataset is balanced by undersampling the majority class (benign) to ensure equal class representation \cite{zhuang2024non}, yielding 424 samples (296 training, 128 test).

\begin{table}[htbp]
\centering
\caption{Experimental Configuration}
\begin{tabular}{|l|l|}
\hline
\textbf{Parameter} & \textbf{Configuration} \\
\hline
\hline
Datasets & MNIST, Fashion-MNIST, Iris, \\
& Breast Cancer \\
\hline
Dataset Sizes & MNIST: 400, Fashion-MNIST: 400, \\ & Iris: 100, Breast Cancer: 424 \\
\hline
Train-Test Split & 70\%-30\% and \\
& $66.\overline{6}$\%-$33.\overline{3}$\% (Breast Cancer) \\
\hline
Reduced Feature Dimension & 4 \\
\hline
Number of Qubits & 4 (angle encoding) \\
\hline
\multirow{2}{*}{Ansatze Layers} & QNN 1: 2 layers  \\
& QNN 2: 4 layers\\
\hline
Training Epochs & QNN 1: 40, QNN 2: 20 \\
\hline
Batch Size & 14 \\
\hline
Learning Rate & 0.02 \\
\hline
Gradient Method & Autograd  \\
\hline
Loss Function & Binary cross-entropy \\
\hline
Number of Trials & 5 (different random seeds) \\
\hline
DAC Resolutions & 2, 4, 6, 8, 10, 12 bits \\
\hline
Temperature Values & 0.5, 1.0, 5.0, 10.0 \\
\hline
\end{tabular}
\label{tab:experimental_configuration}
\vspace{-0.5cm}
\end{table}

\subsection{Data Preprocessing and Feature Encoding}

Data preprocessing varies by dataset complexity. For MNIST and Fashion-MNIST, the original 784-dimensional ($28\times28$) pixel images are reduced to 4 principal components via PCA, capturing the most significant variance. Similarly, the 30-dimensional breast cancer feature set undergoes PCA reduction to 4 components. The Iris dataset, with its native 4 features (sepal length/width, petal length/width), requires no dimensionality reduction and is used directly. All feature sets are normalized to $[-\pi,+\pi]$ to match the periodic domain of quantum rotation gates.

The QNN architectures, illustrated in Figure \ref{fig:methodology_figure}, employ angle encoding to embed classical data into quantum states. Each of the 4 features is encoded via an $R_y(x_i)$ rotation gate applied to qubit $i$, where $x_i \in [-\pi,+\pi]$ denotes the $i$-th feature value.

\subsection{Quantum Neural Network Architectures}

We evaluate two QNN architectures with different expressivity, entangling strategy and parameter counts across all 4 datasets. 

\textbf{QNN 1} employs a compact ansatze architecture with 16 trainable parameters across 2 layers. Each layer applies trainable $R_y$ and $R_z$ rotation gates (2 parameters per qubit) followed by CNOT entangling gates in circular connectivity \cite{sim2019expressibility}, where each qubit connects to its neighbor and the final qubit wraps back to the first. 

\textbf{QNN 2} features 48 trainable parameters across 4 ansatze layers. Each layer applies $R_x$, $R_y$, and $R_z$ gates (3 parameters per qubit), followed by CZ gates in all-to-all connectivity where every qubit pair shares an entangling gate. This architecture provides stronger entanglement and greater representational capacity compared to QNN 1 owing to the higher number of parameters.

For both architectures, classification is performed by measuring the first qubit in the computational basis to obtain the expectation value of the Pauli-$Z$ observable, $\langle\hat{Z}\rangle \in [-1,+1]$. The binary decision rule is: $\langle\hat{Z}\rangle > 0$ predicts class 1, otherwise class 0.

\subsection{Experimental Paradigm 1: Inference with Post-Training Quantization}

We first examine the quality of inference of pre-trained QNNs (trained with infinite precision) when deployed on quantum computers with finite-resolution control electronics. For an $N$-bit DAC, rotation angles are constrained to $2^N$ discrete levels in $[-\pi,+\pi]$ with step size $\Delta = 2\pi/(2^N-1)$. QNN 1 and QNN 2 are trained for 40 and 20 epochs respectively, where the difference reflects QNN 2's faster accuracy convergence due to higher expressivity (48 vs. 16 parameters). Trained parameters and input features are then quantized by rounding to the nearest allowed level for DAC resolutions ranging from 2 to 12 bits, and test accuracy is measured at each resolution.

\begin{figure*}[t]
  \centering
  \begin{subfigure}[b]{0.25\textwidth}
    \centering
    \includegraphics[width=\linewidth]{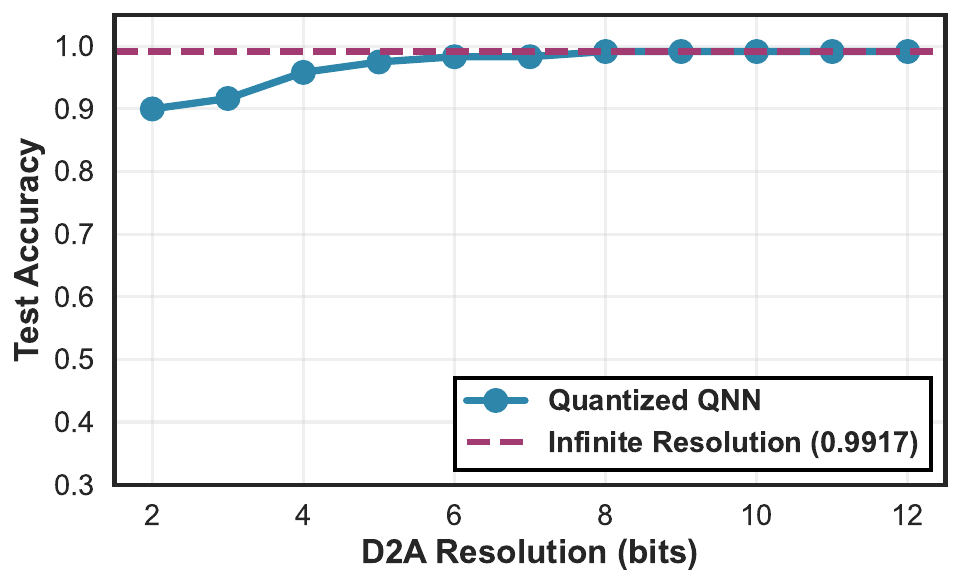}
    \caption{MNIST}
    \label{fig:ptr_mnist}
  \end{subfigure}%
  \hfill
  \begin{subfigure}[b]{0.25\textwidth}
    \centering
    \includegraphics[width=\linewidth]{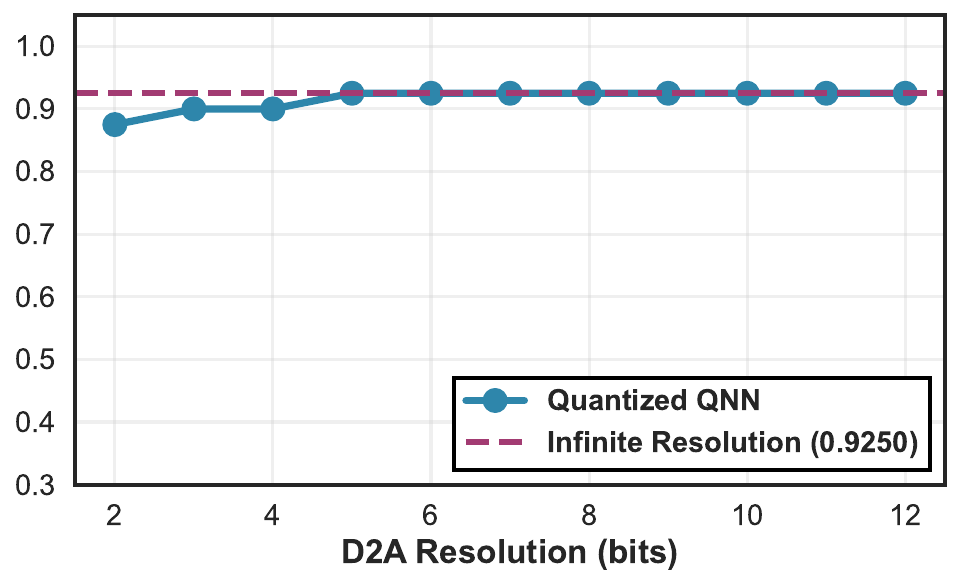}
    \caption{Fashion-MNIST}
    \label{fig:ptr_fmnist}
  \end{subfigure}%
  \hfill
  \begin{subfigure}[b]{0.25\textwidth}
    \centering
    \includegraphics[width=\linewidth]{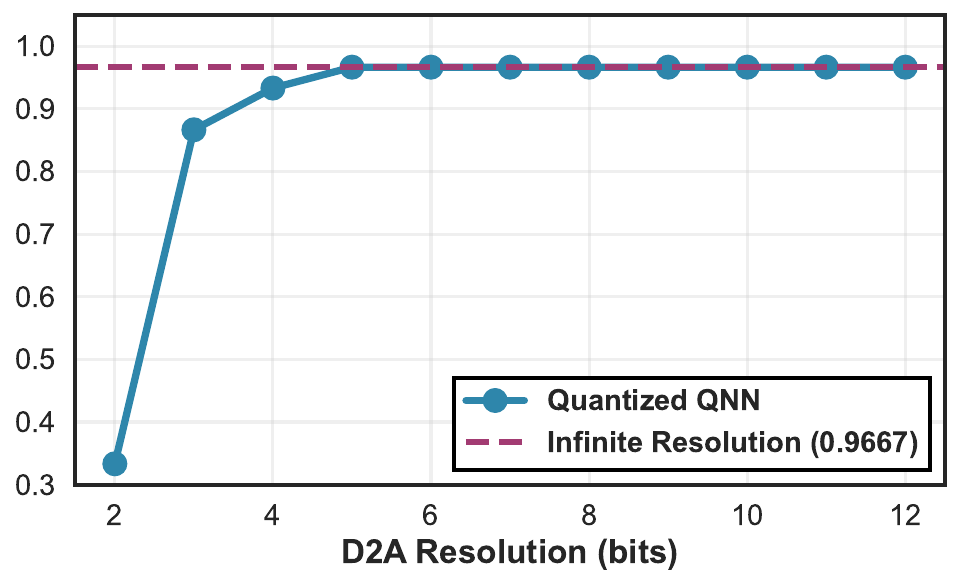}
    \caption{Iris}
    \label{fig:ptr_iris}
  \end{subfigure}%
  \hfill
  \begin{subfigure}[b]{0.25\textwidth}
    \centering
    \includegraphics[width=\linewidth]{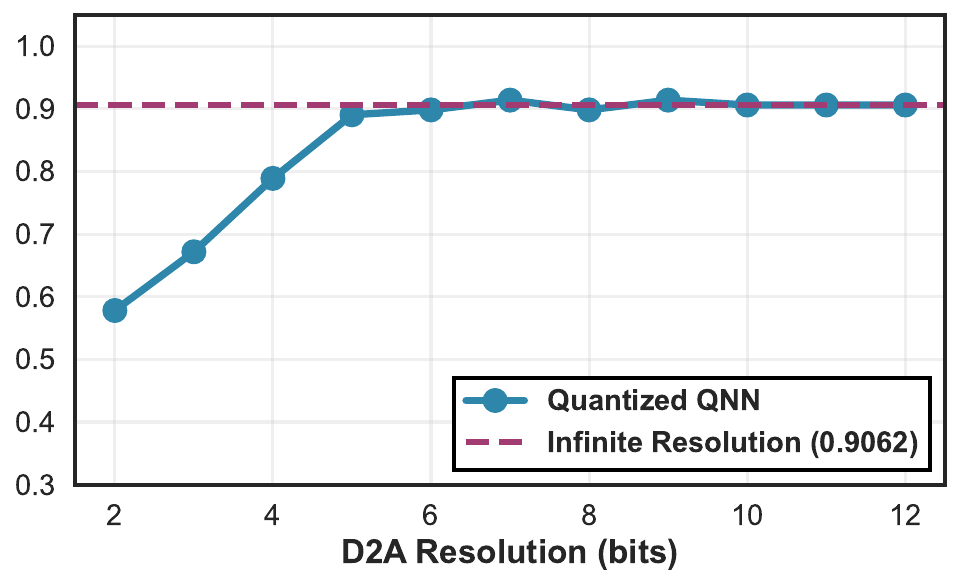}
    \caption{Breast Cancer}
    \label{fig:ptr_bc}
  \end{subfigure}

  \caption{Inference accuracy of pre-trained QNN 1 (infinite precision) as a function of D2A resolution upon deployment on a quantum computer with limited resolution D2As. Dotted line shows the test accuracy of infinite resolution QNN 1.}
  \label{fig:ptr_qnn1}
\end{figure*}


\begin{figure*}[t]
  \centering
  \begin{subfigure}[b]{0.25\textwidth}
    \centering
    \includegraphics[width=\linewidth]{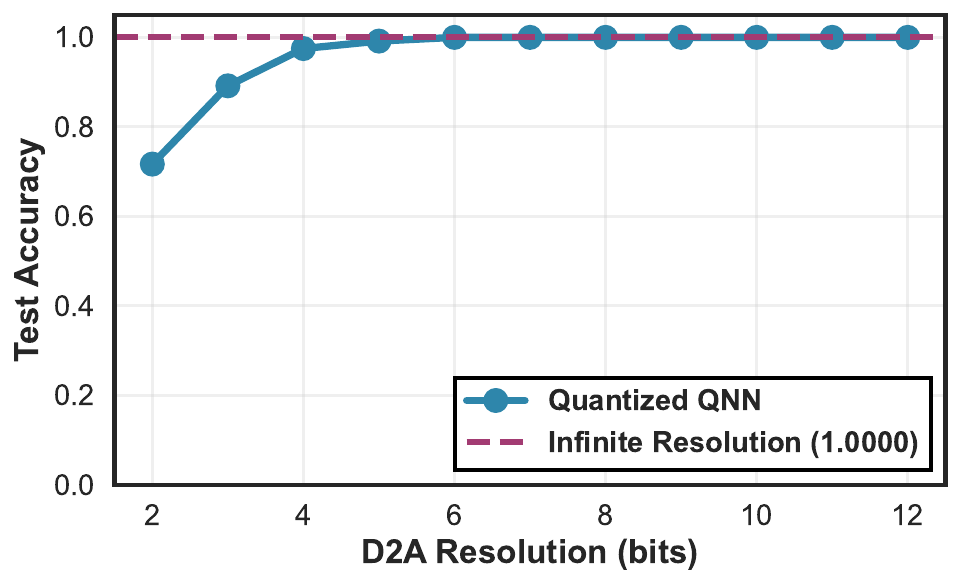}
    \caption{MNIST}
    \label{fig:ptr_mnist2}
  \end{subfigure}%
  \hfill
  \begin{subfigure}[b]{0.25\textwidth}
    \centering
    \includegraphics[width=\linewidth]{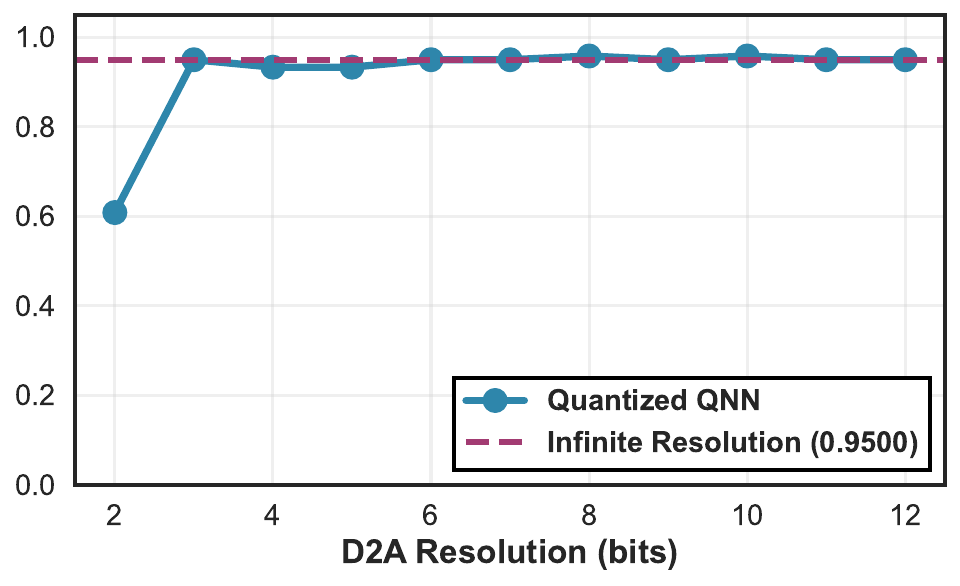}
    \caption{Fashion-MNIST}
    \label{fig:ptr_fmnist2}
  \end{subfigure}%
  \hfill
  \begin{subfigure}[b]{0.25\textwidth}
    \centering
    \includegraphics[width=\linewidth]{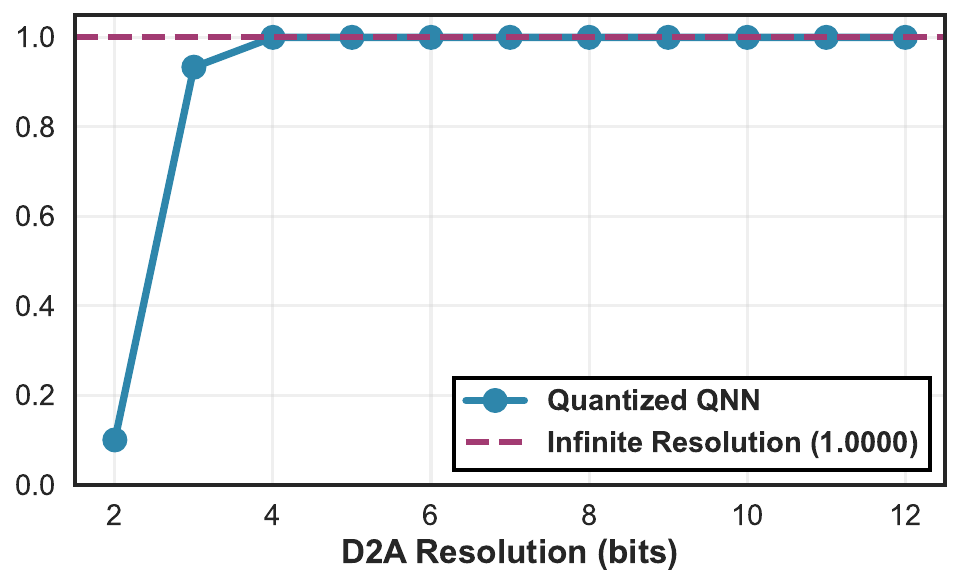}
    \caption{Iris}
    \label{fig:ptr_iris2}
  \end{subfigure}%
  \hfill
  \begin{subfigure}[b]{0.25\textwidth}
    \centering
    \includegraphics[width=\linewidth]{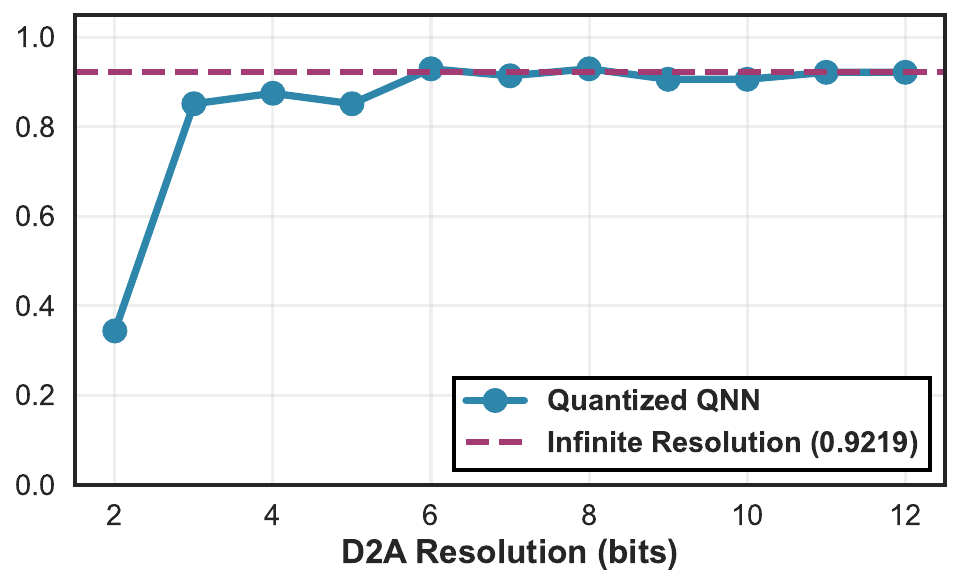}
    \caption{Breast Cancer}
    \label{fig:ptr_bc2}
  \end{subfigure}

  \caption{Inference accuracy of pre-trained QNN 2 (infinite precision) as a function of D2A resolution upon deployment on a quantum computer with limited resolution D2As. Dotted line shows the test accuracy of infinite resolution QNN 2.}
  \label{fig:ptr_qnn2}
\end{figure*}

Throughout this paper, ``infinite precision'' denotes the baseline case where parameters use standard 32-bit floating-point (FP32) arithmetic, unconstrained by DAC quantization. While not mathematically infinite, FP32 provides approximately 7 decimal digits of precision, which is effectively unconstrained relative to the discrete $N$-bit quantization levels studied here.

\begin{figure*}[t]
  \centering
  \begin{subfigure}[b]{0.49\textwidth}
    \centering
    \includegraphics[width=\linewidth]{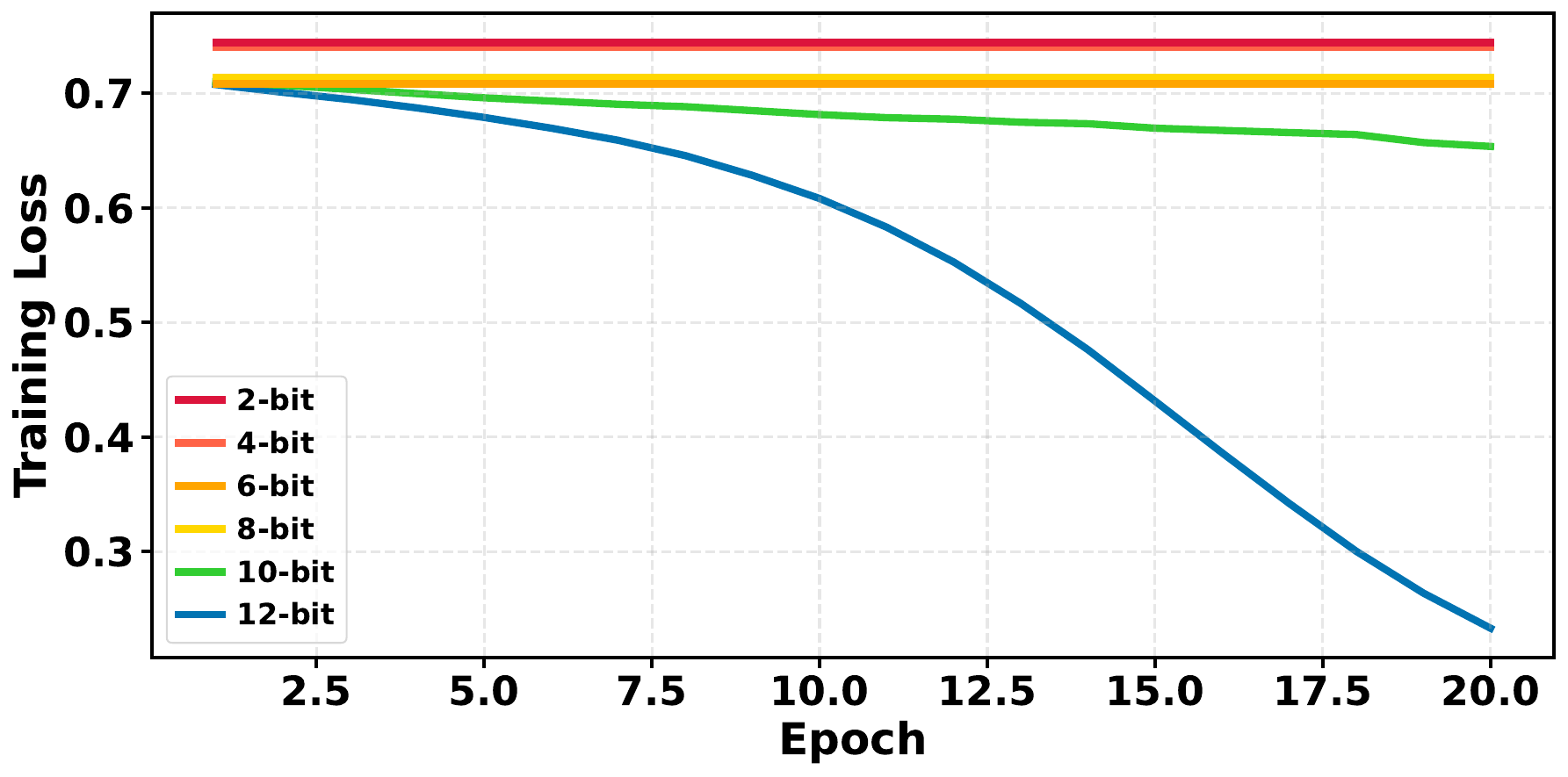}
    \caption{Training curves for deterministic parameter updates.}
    \label{fig:deadlock_a}
  \end{subfigure}%
  \hfill
  \begin{subfigure}[b]{0.49\textwidth}
    \centering
    \includegraphics[width=\linewidth]{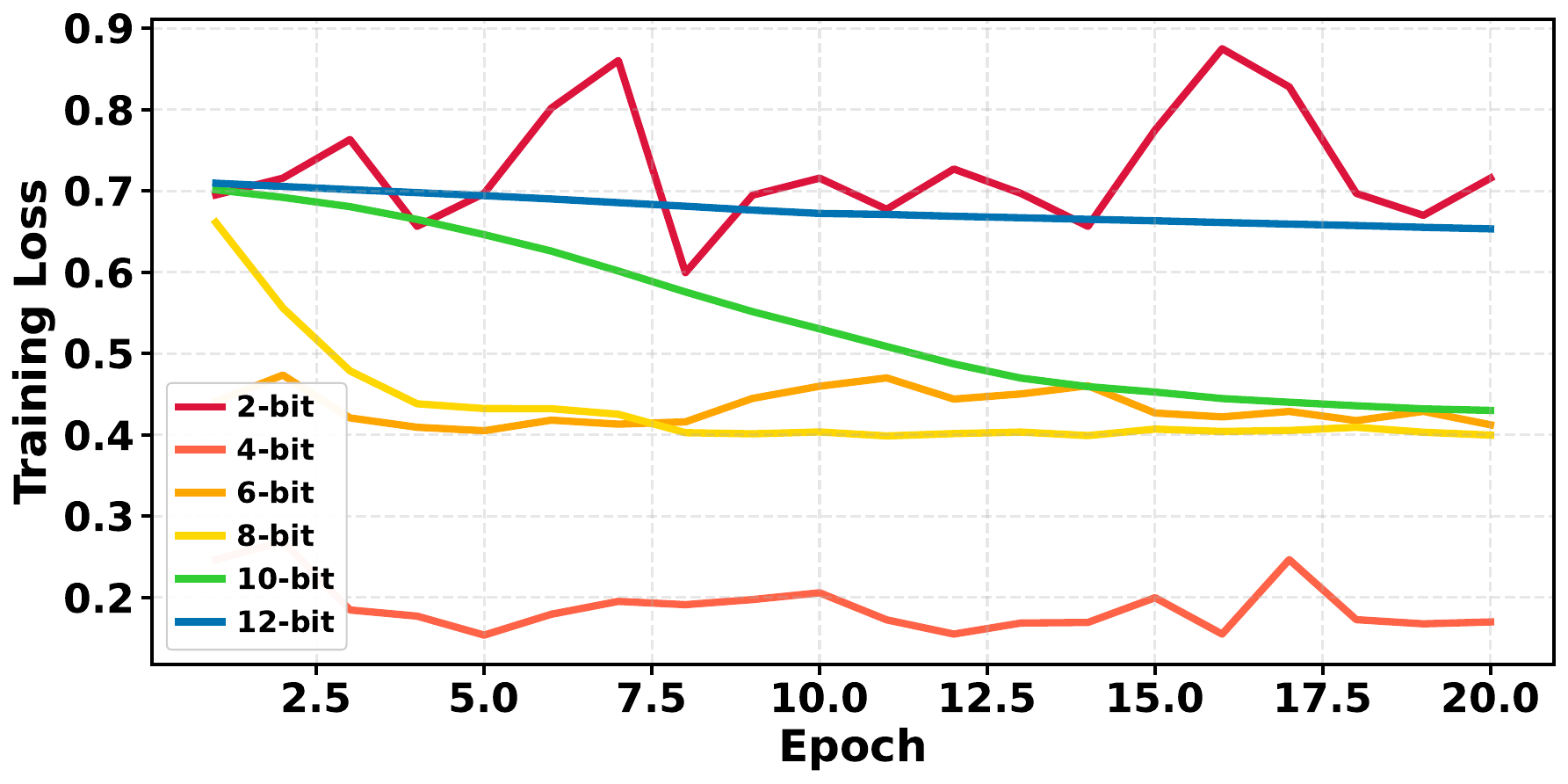}
    \caption{Training curves with stochastic parameter updates.}
    \label{fig:deadlock_b}
  \end{subfigure}
  \vspace{-0.1cm}
  \caption{Training loss vs epochs (single run) for all DAC resolutions with deterministic and stochastic ($T=1.0$) parameter update (QNN 1: MNIST).}
  \vspace{-0.5cm}
  \label{fig:combined_deadlock}
\end{figure*}

\subsection{Experimental Paradigm 2: Training with Quantization}

We next investigate the training of quantized QNNs, where both QNNs are trained with quantization constraints enforced throughout the learning process, and compare performance against infinite-precision baselines. During training, parameters are constrained to discrete $N$-bit values after each gradient update: $\theta \leftarrow \text{quantize}(\theta - \eta \nabla_{\theta}\mathcal{L})$, where $\eta=0.02$ is the learning rate and $\mathcal{L}$ is the binary cross-entropy loss. Parameters are rounded to the nearest quantized level after each gradient step to ensure parameters remain at allowed values based on DAC resolution throughout training, faithfully simulating the constraints of finite-resolution control electronics.

\subsubsection{The Gradient Deadlock Problem}

When the parameter update magnitude (gradient magnitude scaled by learning rate) becomes smaller than half the quantization step size, i.e. $|\eta \nabla_\theta \mathcal{L}| < \Delta/2$, deterministic rounding consistently returns parameters to their current quantized values, preventing any update. This \textit{gradient deadlock} halts learning entirely. The phenomenon is particularly severe at low resolutions where $\Delta$ is large, and during later training epochs when gradients naturally become smaller as the optimizer approaches minima.

\subsection{Proposed Solution: Temperature-Controlled Stochastic Quantization}

To overcome gradient deadlock, we introduce stochastic parameter updates controlled by a temperature hyperparameter $T$. Rather than deterministically rounding to the nearest quantization level, we probabilistically decide whether to jump to adjacent levels based on:
\begin{equation}
P(\theta_{\text{next}}) = \frac{1}{1 + \exp(-d/T)}
\end{equation}
where $d$ is the normalized distance from the continuous update to the midpoint between current and next quantization levels:
\begin{equation}
d = \frac{2}{\Delta}(\theta - \eta \nabla_\theta \mathcal{L} - m)
\end{equation}
where $m$ denotes the midpoint between the two quantization levels that enclose the continuous update value. The sigmoid function ensures parameters favor the level closest to the continuous update $\theta - \eta \nabla_\theta \mathcal{L}$ while allowing exploration through controlled stochasticity. Higher temperature $T$ increases randomness and $T \to 0$ recovers deterministic rounding. Note that this stochastic quantization addresses gradient deadlock due to hardware-imposed discrete parameter spaces, distinct from stochastic optimization methods like simulated annealing \cite{10.5555/329748.329752, 701179} or stochastic gradient descent \cite{Rumelhart1986} which introduce noise for exploring continuous landscapes \cite{Goodfellow-et-al-2016}.

\subsection{Experimental Protocol}

We systematically evaluate resolutions of 2, 4, 6, 8, 10, and 12 bits, with temperature values 0.5, 1.0, 5.0, and 10.0 for each resolution. Each configuration is trained for 5 independent runs with different random initialization seeds to ensure statistical robustness. Performance is evaluated using average test accuracy across trials. All experiments were conducted using PennyLane's \cite{bergholm2018pennylane} \texttt{lightning.qubit} high-performance simulator. Training hyperparameters are listed in Table~\ref{tab:experimental_configuration}.

\begin{figure*}[t]
  \centering
  \begin{subfigure}[b]{0.49\textwidth}
    \centering
    \includegraphics[width=\linewidth]{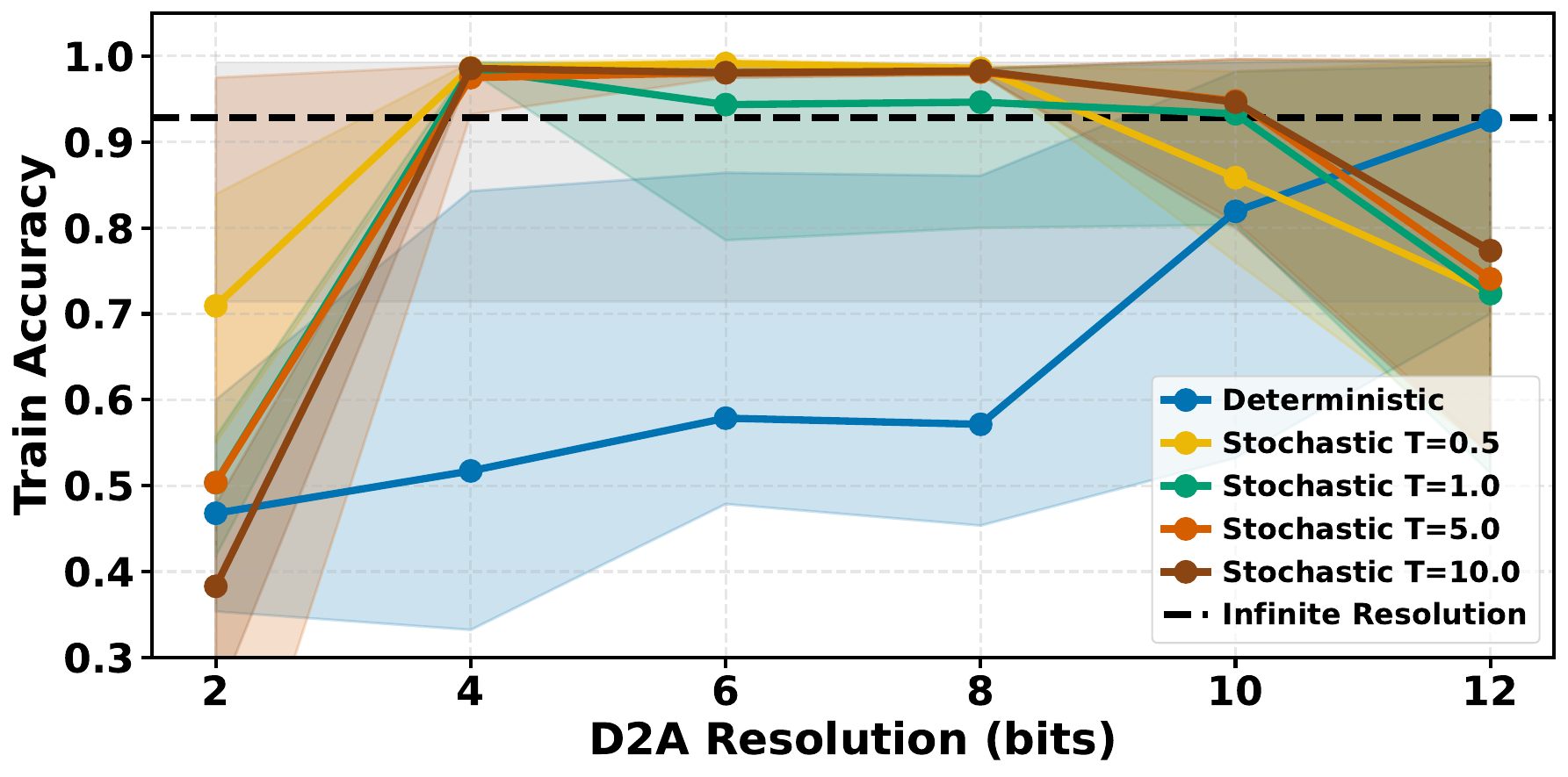}
    \caption{Training accuracy vs Resolution (QNN 1: MNIST)}
    \label{fig:acc_a_mnist}
  \end{subfigure}%
  \hfill
  \begin{subfigure}[b]{0.49\textwidth}
    \centering
    \includegraphics[width=\linewidth]{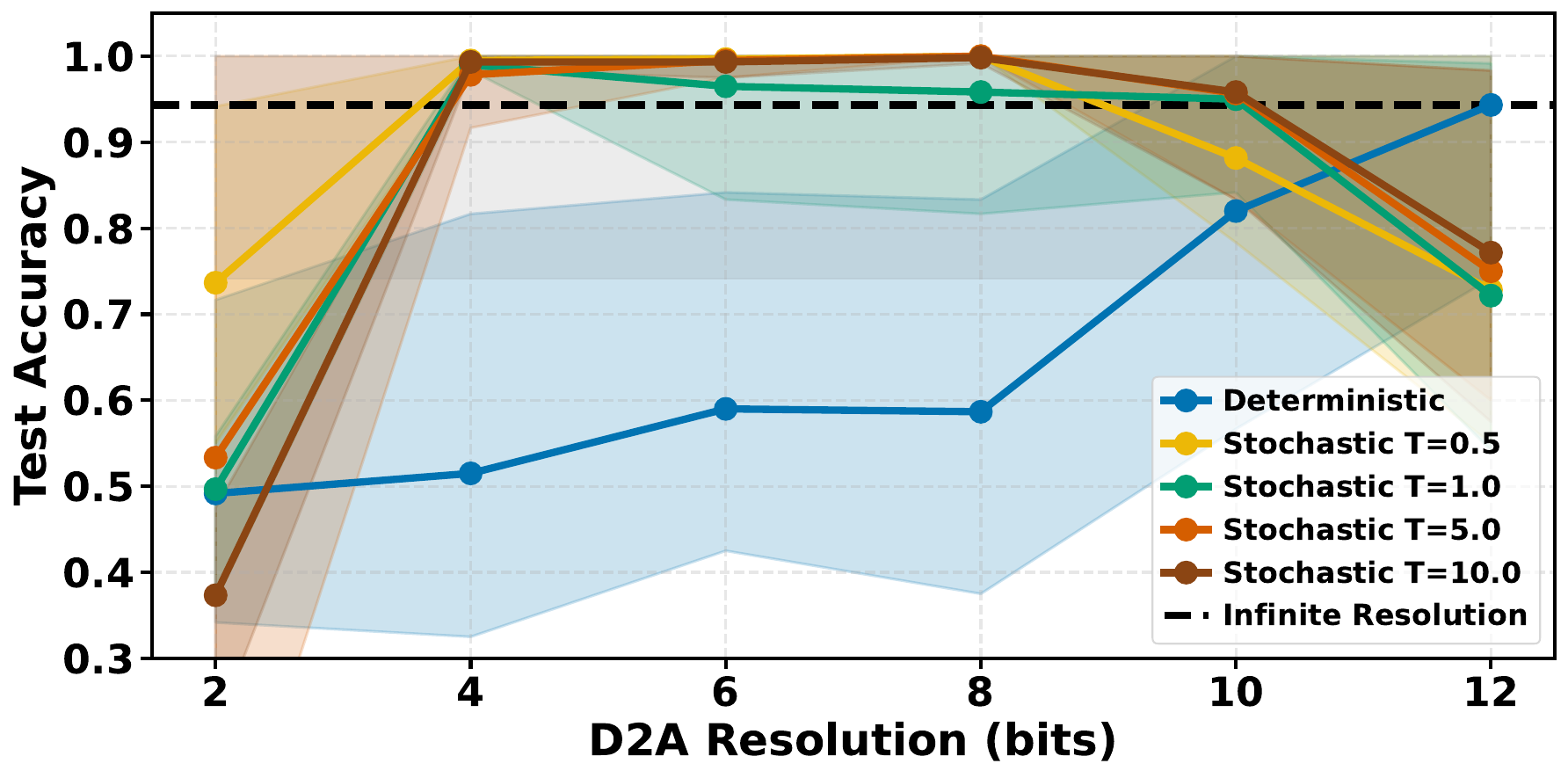}
    \caption{Test accuracy vs Resolution (QNN 1: MNIST).}
    \label{fig:acc_b_mnist}
  \end{subfigure}
  \vspace{-0.1cm}
  \caption{Average train/ test accuracy vs DAC (D2A) resolution on MNIST dataset (QNN 1), for deterministic and stochastic quantization strategies. Shaded regions show variance across 5 trials.}
  \label{fig:train_test_accuracy_mnist}
\end{figure*}

\begin{figure*}[t]
  \centering
  \begin{subfigure}[b]{0.49\textwidth}
    \centering
    \includegraphics[width=\linewidth]{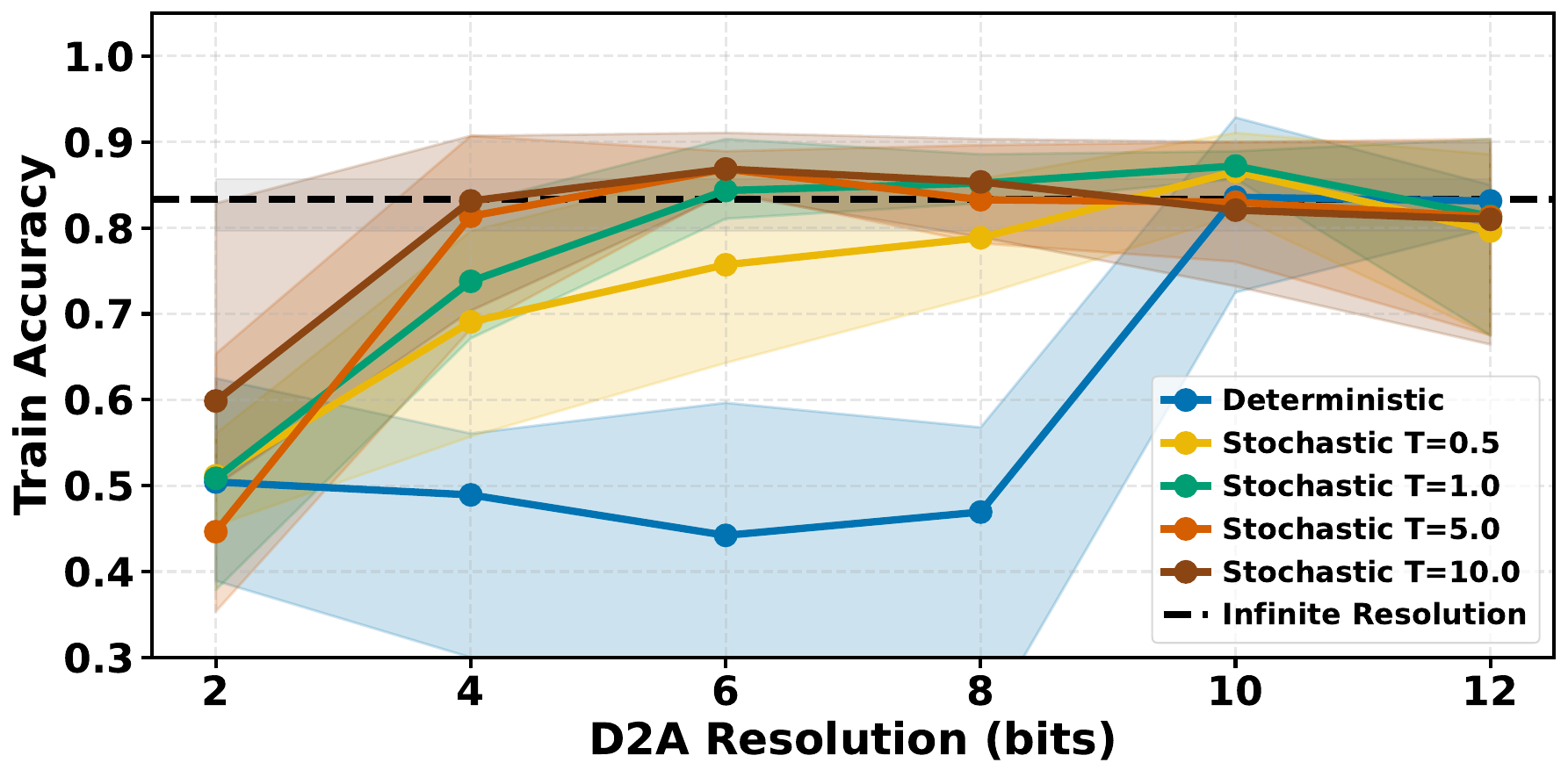}
    \caption{Training accuracy vs Resolution (QNN 1: Fashion-MNIST).}
    \label{fig:acc_a_fmnist}
  \end{subfigure}%
  \hfill
  \begin{subfigure}[b]{0.49\textwidth}
    \centering
    \includegraphics[width=\linewidth]{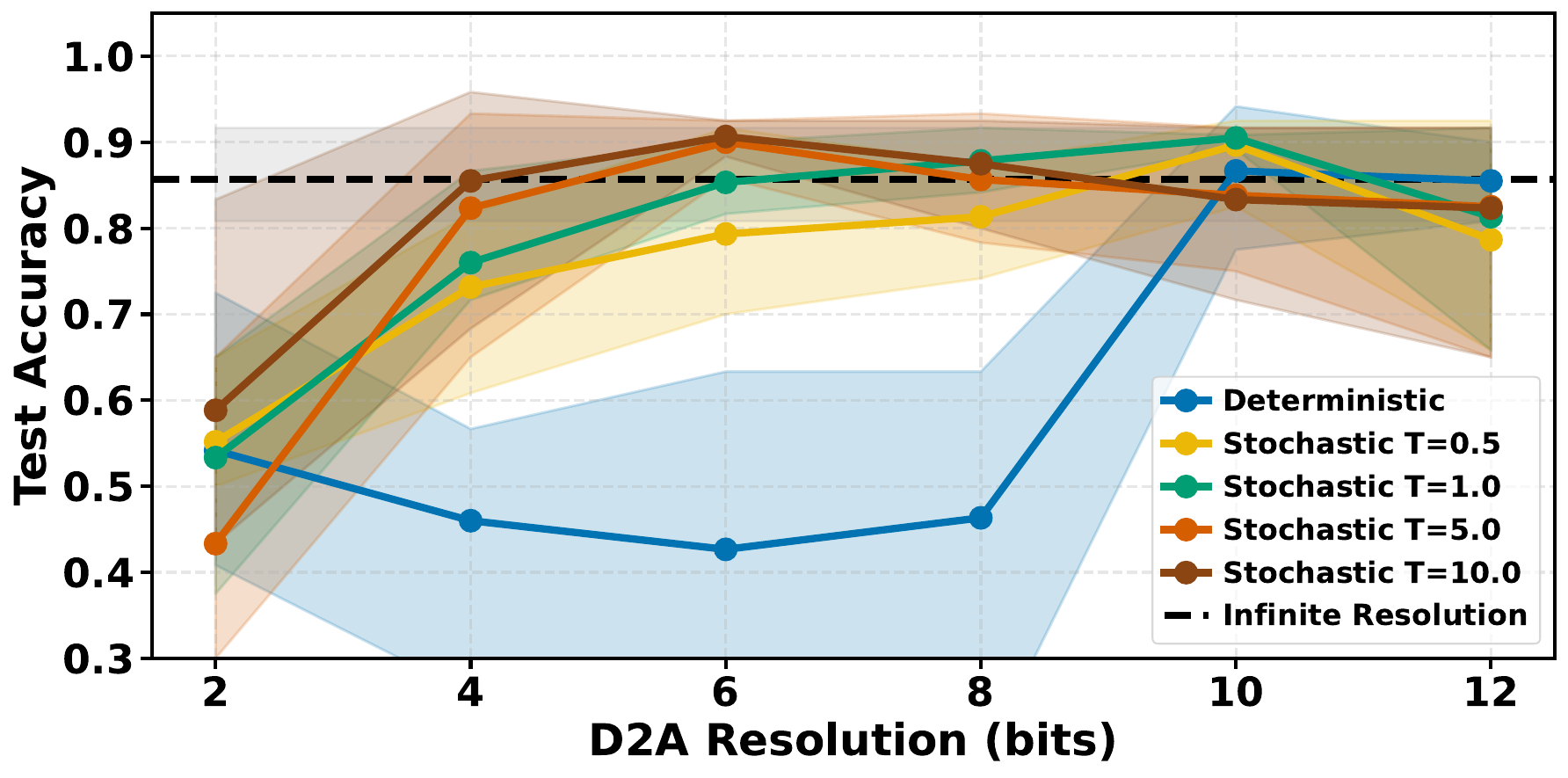}
    \caption{Test accuracy vs Resolution (QNN 1: Fashion-MNIST).}
    \label{fig:acc_b_fmnist}
  \end{subfigure}
  \vspace{-0.1cm}
  \caption{Average train/ test accuracy vs DAC (D2A) resolution, on Fashion-MNIST dataset (QNN 1), for deterministic and stochastic quantization strategies. Shaded regions show variance across 5 trials.}
  \label{fig:train_test_accuracy_fmnist}
\end{figure*}

\begin{figure*}[t]
  \centering
  \begin{subfigure}[b]{0.49\textwidth}
    \centering
    \includegraphics[width=\linewidth]{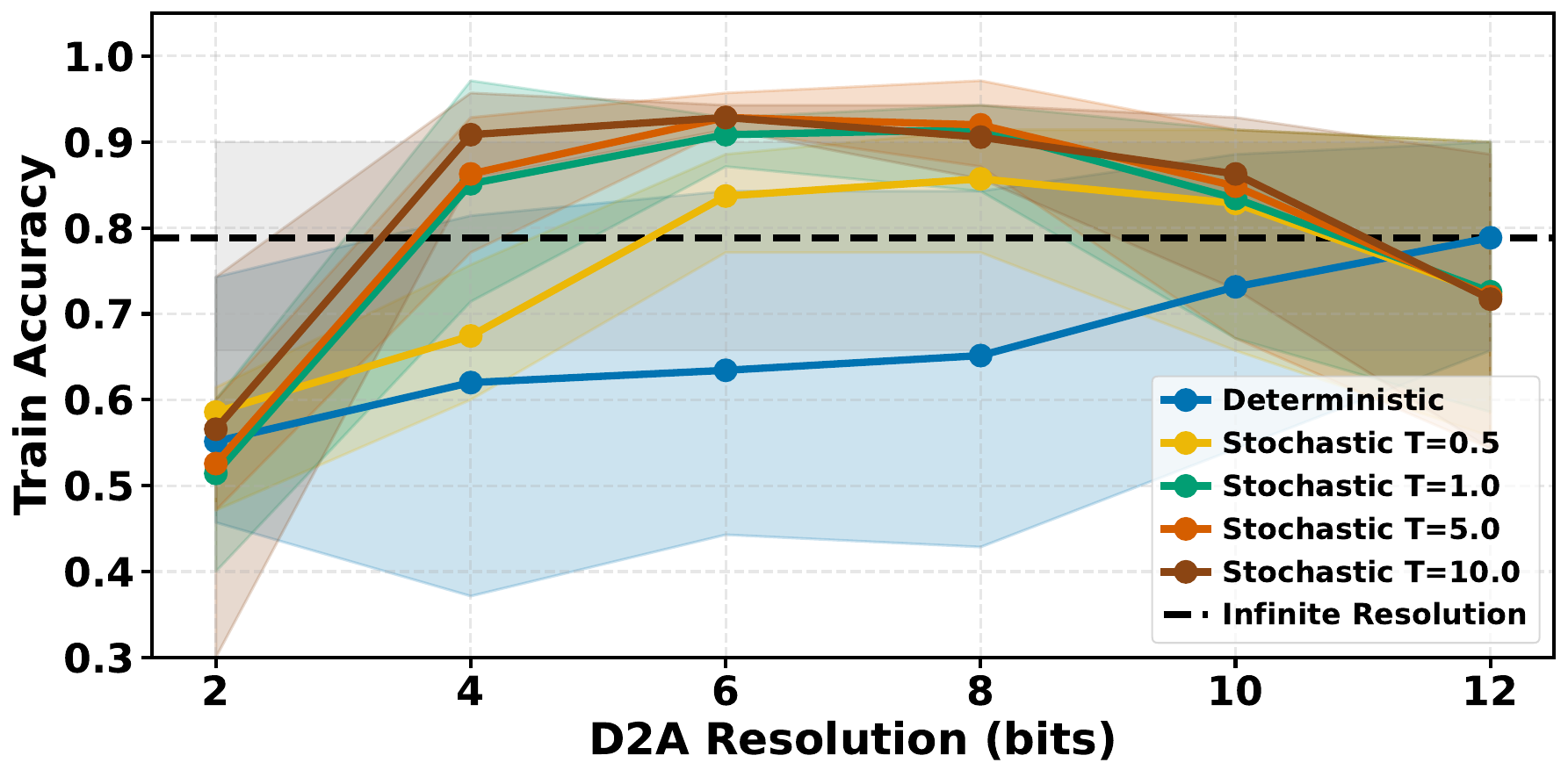}
    \caption{Training accuracy vs Resolution (QNN 1: Iris).}
    \label{fig:acc_a_iris}
  \end{subfigure}%
  \hfill
  \begin{subfigure}[b]{0.49\textwidth}
    \centering
    \includegraphics[width=\linewidth]{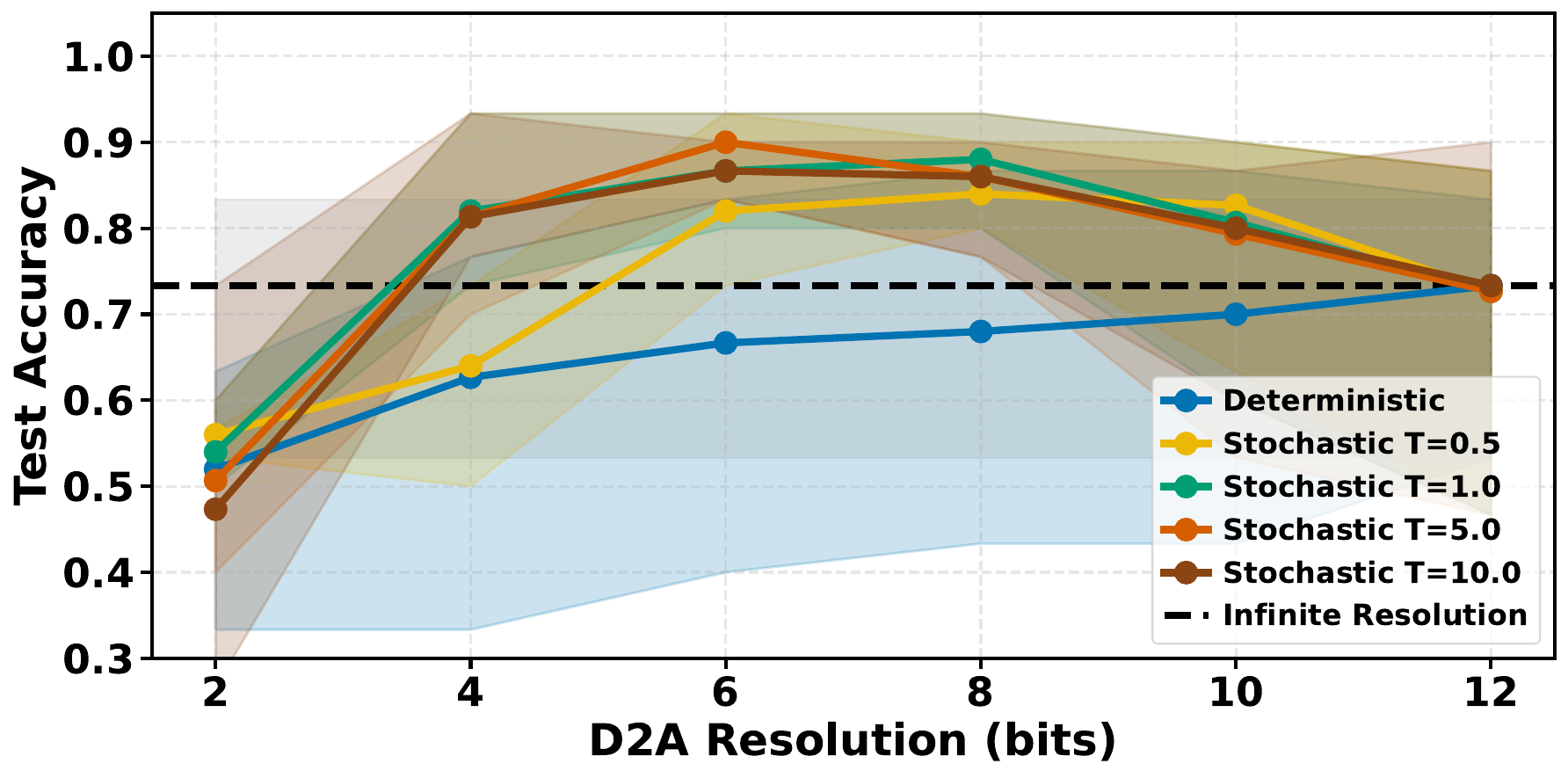}
    \caption{Test accuracy vs Resolution (QNN 1: Iris).}
    \label{fig:acc_b_iris}
  \end{subfigure}
  \vspace{-0.1cm}
  \caption{Average train/ test accuracy vs DAC (D2A) resolution, on Iris dataset (QNN 1), for deterministic and stochastic quantization strategies. Shaded regions show variance across 5 trials.}
  \label{fig:train_test_accuracy_iris}
\end{figure*}

\begin{figure*}[t]
  \centering
  \begin{subfigure}[b]{0.49\textwidth}
    \centering
    \includegraphics[width=\linewidth]{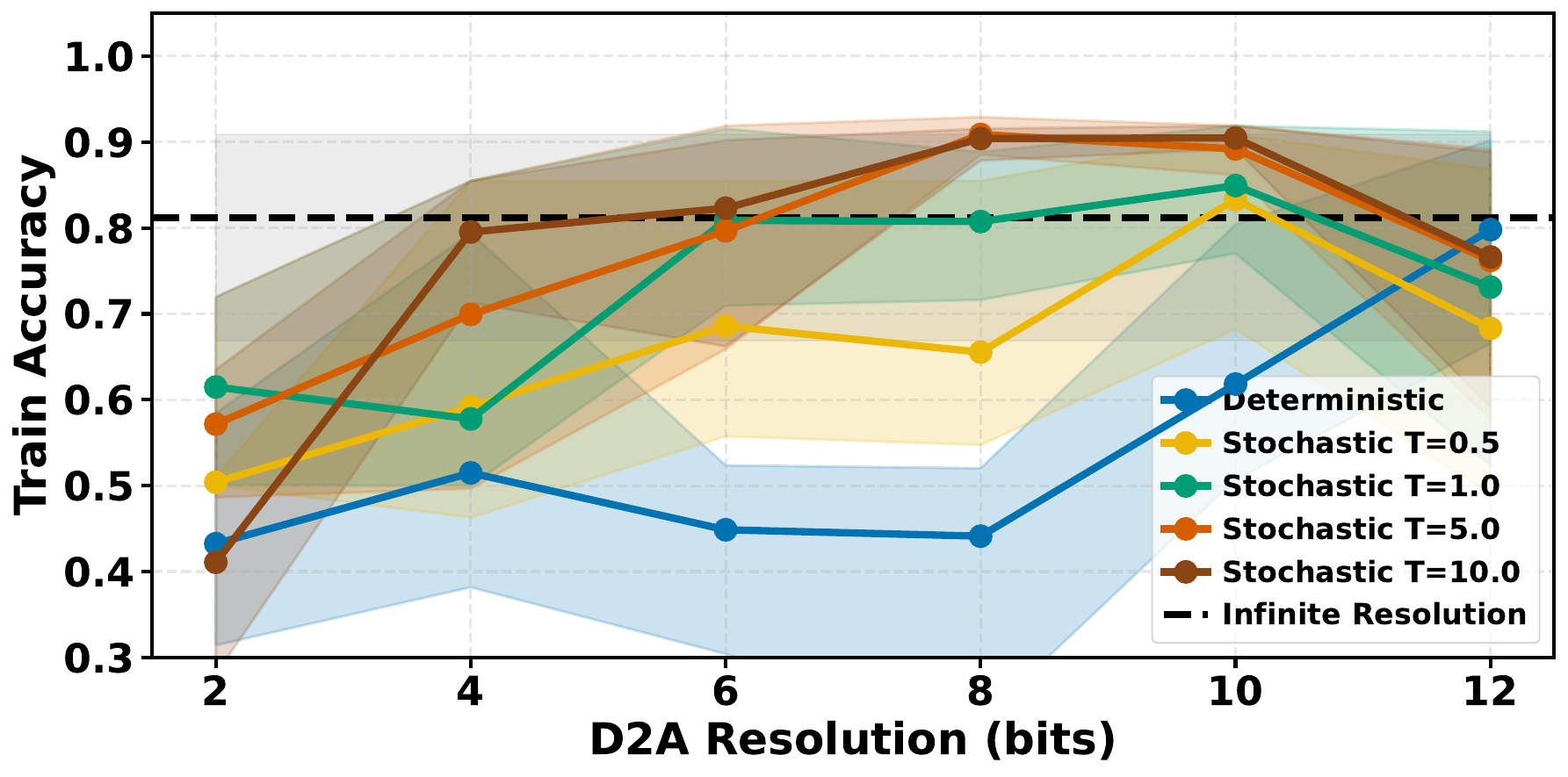}
    \caption{Training accuracy vs Resolution (QNN 1: Breast Cancer).}
    \label{fig:acc_a_bc}
  \end{subfigure}%
  \hfill
  \begin{subfigure}[b]{0.49\textwidth}
    \centering
    \includegraphics[width=\linewidth]{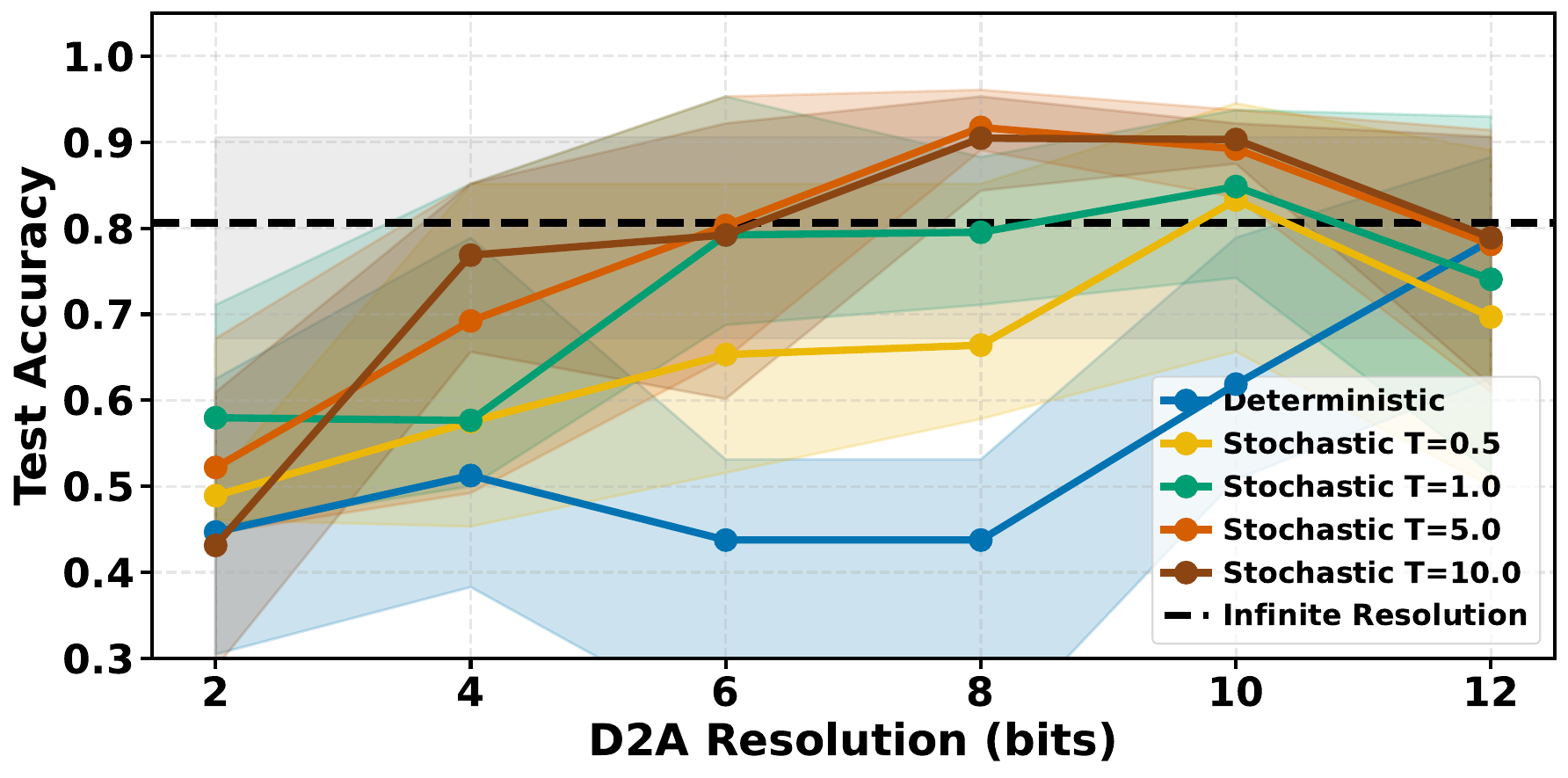}
    \caption{Test accuracy vs Resolution (QNN 1: Breast Cancer).}
    \label{fig:acc_b_bc}
  \end{subfigure}
  \vspace{-0.1cm}
  \caption{Average train/ test accuracy vs DAC (D2A) resolution, on breast cancer dataset (QNN 1), for deterministic and stochastic quantization strategies. Shaded regions show variance across 5 trials.}
  \label{fig:train_test_accuracy_bc}
\end{figure*}

\subsection{Gradient Computation: Autograd vs. Parameter-Shift}

We employ automatic differentiation (autograd) for gradient computation, a standard practice in simulation-based QML \cite{bowles2024better, bergholm2018pennylane} that provides exact gradients through backpropagation and is computationally more efficient than the parameter-shift rule on classical simulators. Real quantum devices, particularly large-scale systems beyond classical simulation capacity, require the parameter-shift rule: $\nabla_{\theta}\mathcal{L} = \frac{1}{2}[\mathcal{L}(\theta+\pi/2) - \mathcal{L}(\theta-\pi/2)]$, which evaluates quantum circuits at shifted angles $\theta \pm \pi/2$ \cite{schuld2019evaluating}. However, under quantization, these shifted angles may not align with allowed discrete values and require rounding, introducing gradient approximation errors on real hardware, a compounding issue particularly severe at low DAC resolutions where quantization step sizes are large. Our simulation approach avoids this gradient-level quantization problem while maintaining parameters at discrete $N$-bit values throughout training, while also enabling efficient systematic exploration of our large experimental space (1240 total runs: 155 runs each for 4 datasets across both QNNs).  We acknowledge that our findings may not fully capture training dynamics on large-scale quantum devices at very low resolutions, where gradient shifts due to quantization become significant. Future studies should validate these results using parameter-shift implementations on simulators and real hardware.
\vspace{-0.15cm}


\begin{figure*}[t]
  \centering
  \begin{subfigure}[b]{0.49\textwidth}
    \centering
    \includegraphics[width=\linewidth]{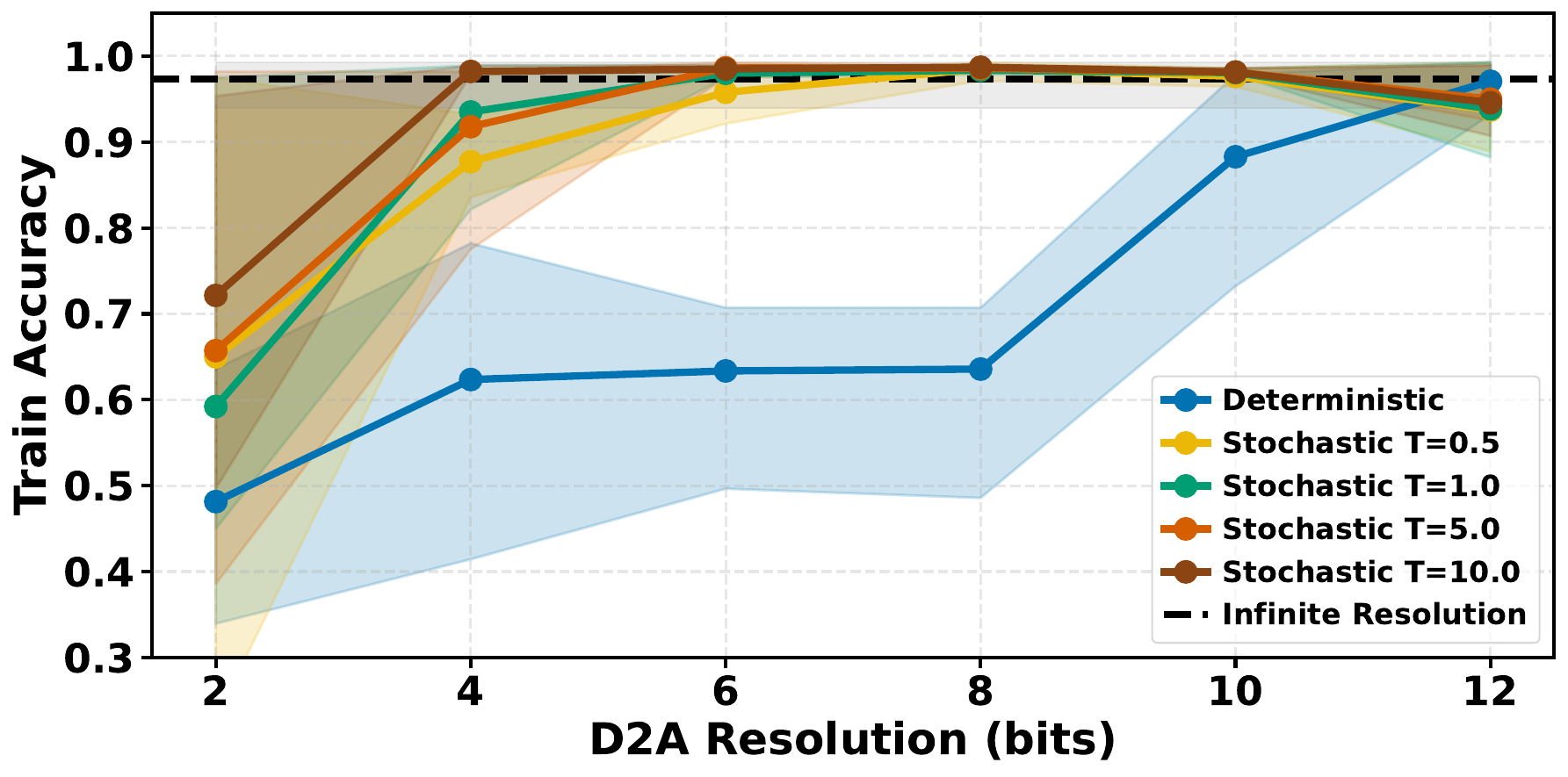}
    \caption{Training accuracy vs Resolution (QNN 2: MNIST)}
    \label{fig:acc_a_mnist2}
  \end{subfigure}%
  \hfill
  \begin{subfigure}[b]{0.49\textwidth}
    \centering
    \includegraphics[width=\linewidth]{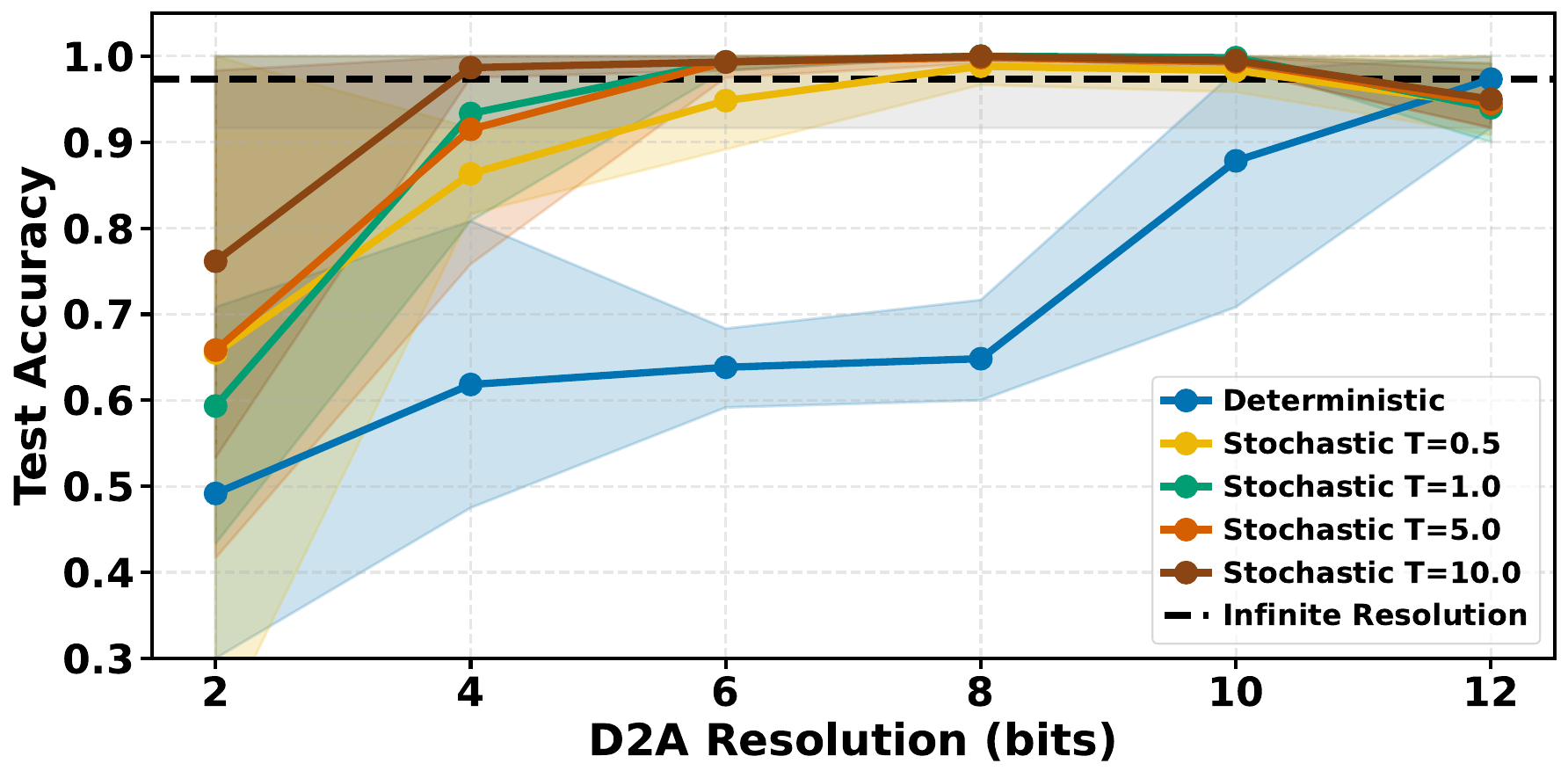}
    \caption{Test accuracy vs Resolution (QNN 2: MNIST).}
    \label{fig:acc_b_mnist2}
  \end{subfigure}
  \vspace{-0.1cm}
  \caption{Average train/ test accuracy vs DAC (D2A) resolution on MNIST dataset (QNN 2), for deterministic and stochastic quantization strategies. Shaded regions show variance across 5 trials.}
  \label{fig:train_test_accuracy_mnist2}
\end{figure*}

\begin{figure*}[t]
  \centering
  \begin{subfigure}[b]{0.49\textwidth}
    \centering
    \includegraphics[width=\linewidth]{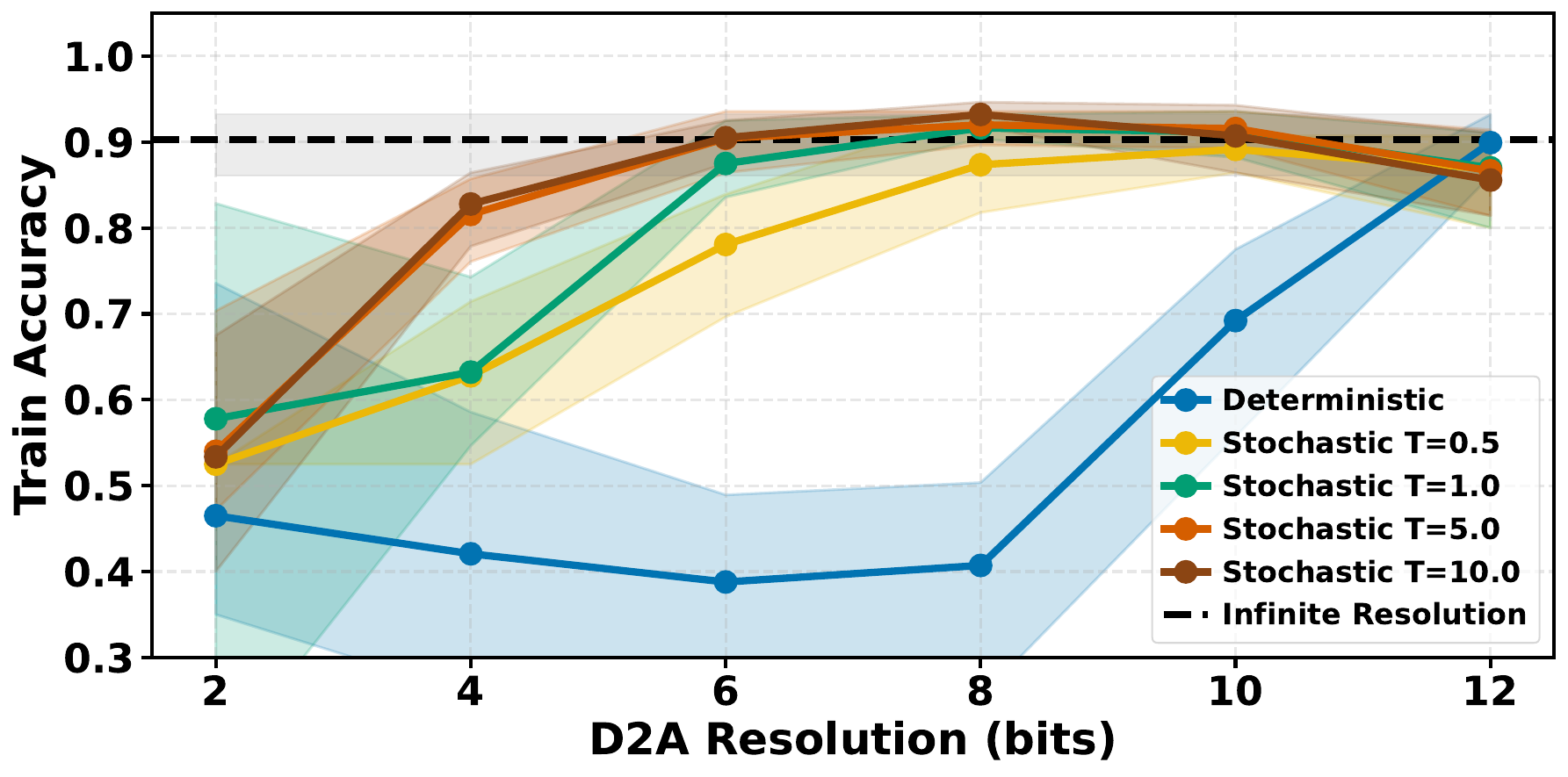}
    \caption{Training accuracy vs Resolution (QNN 2: Fashion-MNIST).}
    \label{fig:acc_a_fmnist2}
  \end{subfigure}%
  \hfill
  \begin{subfigure}[b]{0.49\textwidth}
    \centering
    \includegraphics[width=\linewidth]{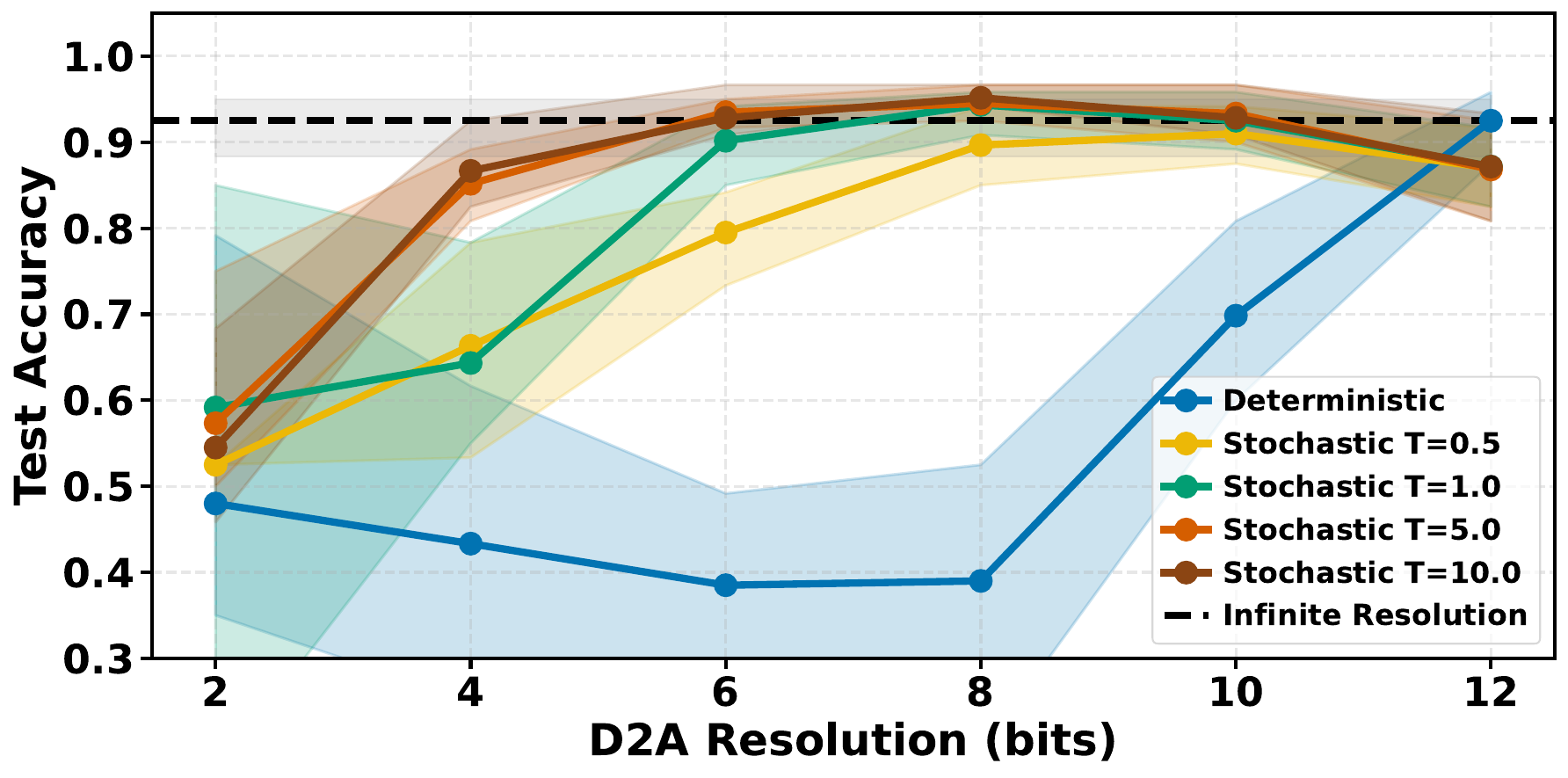}
    \caption{Test accuracy vs Resolution (QNN 2: Fashion-MNIST).}
    \label{fig:acc_b_fmnist2}
  \end{subfigure}
  \vspace{-0.1cm}
  \caption{Average train/ test accuracy vs DAC (D2A) resolution on Fashion- MNIST dataset (QNN 2), for deterministic and stochastic quantization strategies. Shaded regions show variance across 5 trials.}
  \label{fig:train_test_accuracy_fmnist2}
\end{figure*}

\begin{figure*}[t]
  \centering
  \begin{subfigure}[b]{0.49\textwidth}
    \centering
    \includegraphics[width=\linewidth]{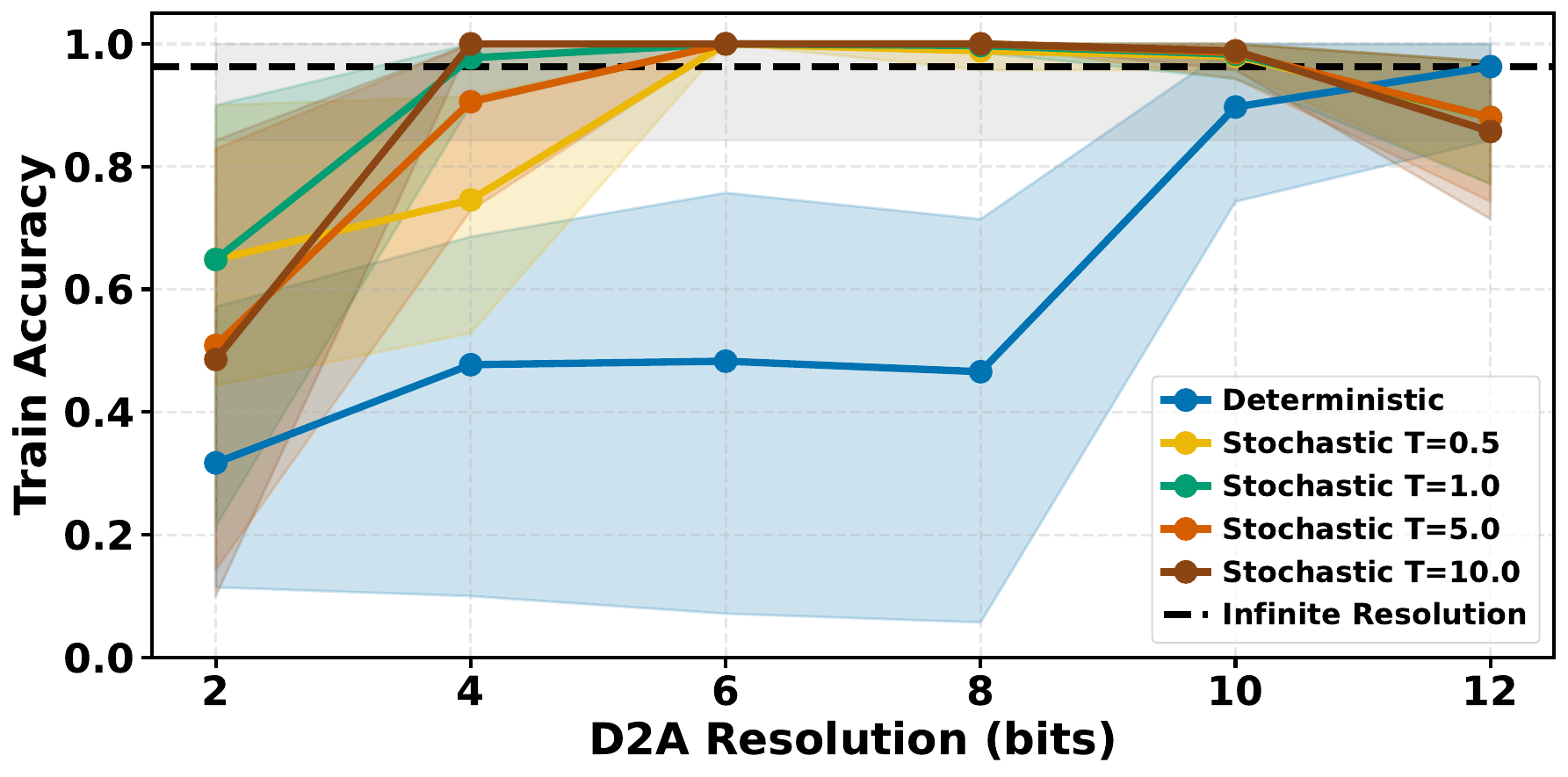}
    \caption{Training accuracy vs Resolution (QNN 2: Iris).}
    \label{fig:acc_a_iris2}
  \end{subfigure}%
  \hfill
  \begin{subfigure}[b]{0.49\textwidth}
    \centering
    \includegraphics[width=\linewidth]{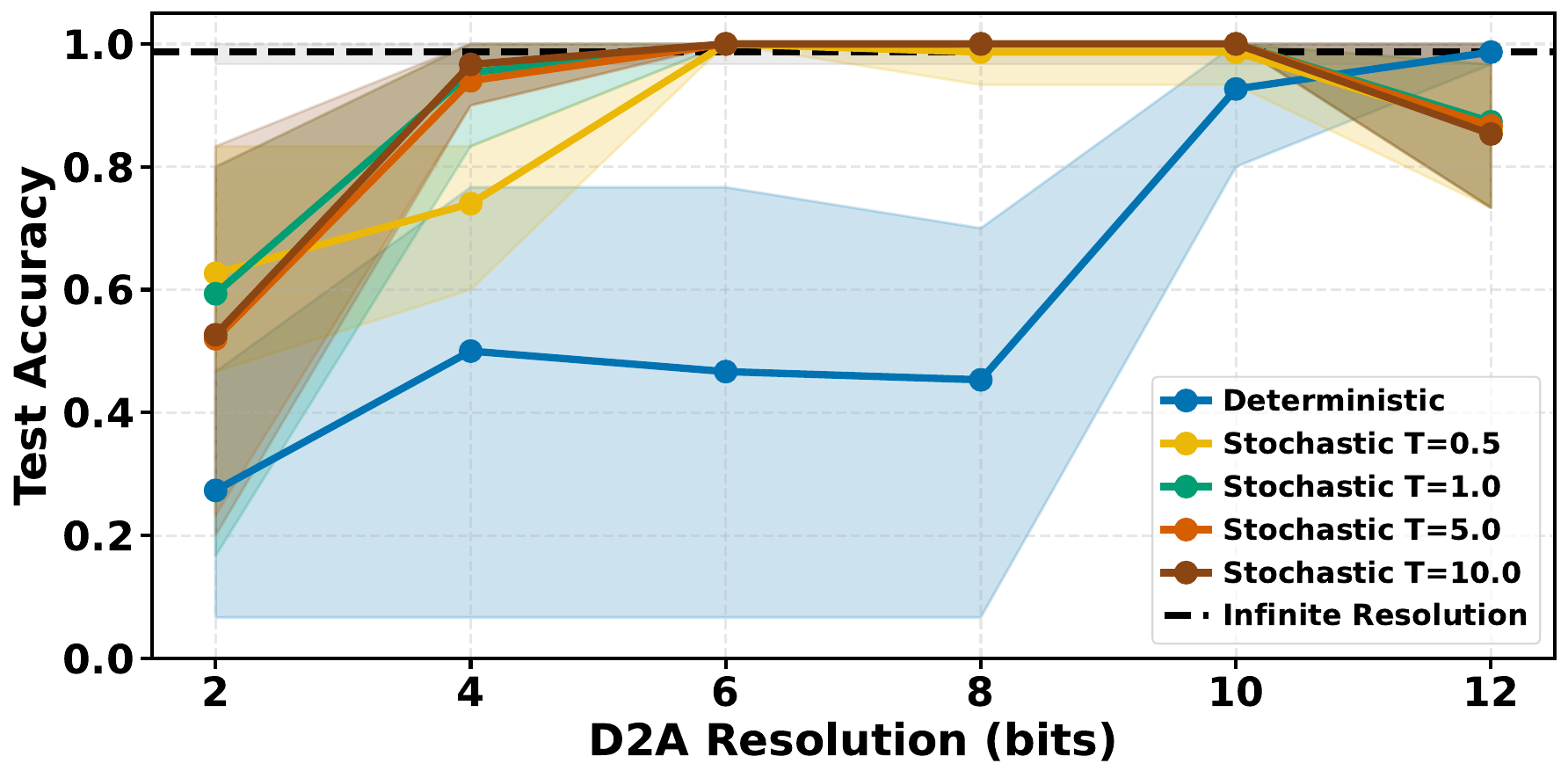}
    \caption{Test accuracy vs Resolution (QNN 2: Iris).}
    \label{fig:acc_b_iris2}
  \end{subfigure}
  \vspace{-0.1cm}
  \caption{Average train/ test accuracy vs DAC (D2A) resolution on Iris dataset (QNN 2), for deterministic and stochastic quantization strategies. Shaded regions show variance across 5 trials.}
  \label{fig:train_test_accuracy_iris2}
\end{figure*}

\begin{figure*}[t]
  \centering
  \begin{subfigure}[b]{0.49\textwidth}
    \centering
    \includegraphics[width=\linewidth]{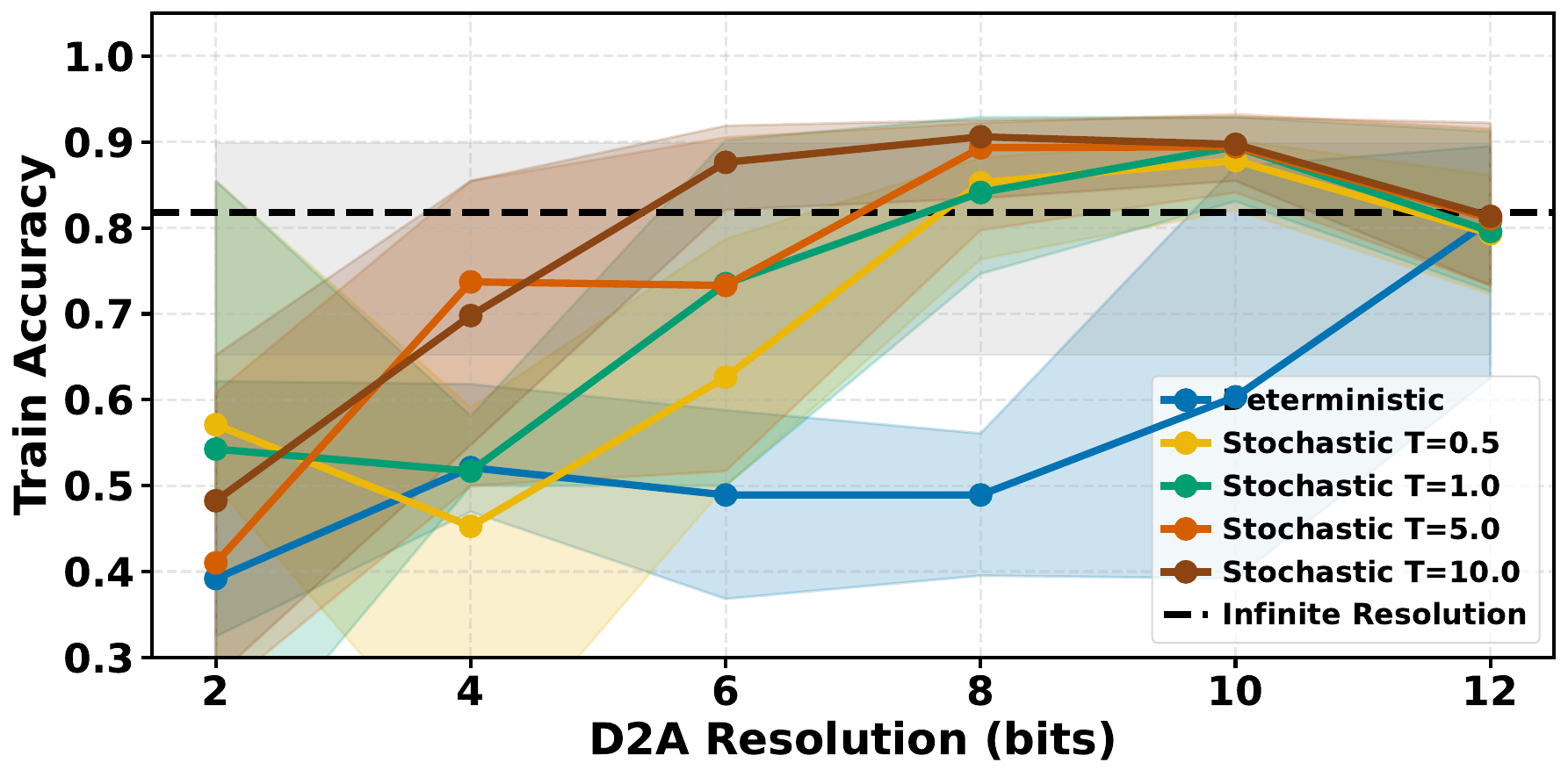}
    \caption{Training accuracy vs Resolution (QNN 2: Breast Cancer).}
    \label{fig:acc_a_bc2}
  \end{subfigure}%
  \hfill
  \begin{subfigure}[b]{0.49\textwidth}
    \centering
    \includegraphics[width=\linewidth]{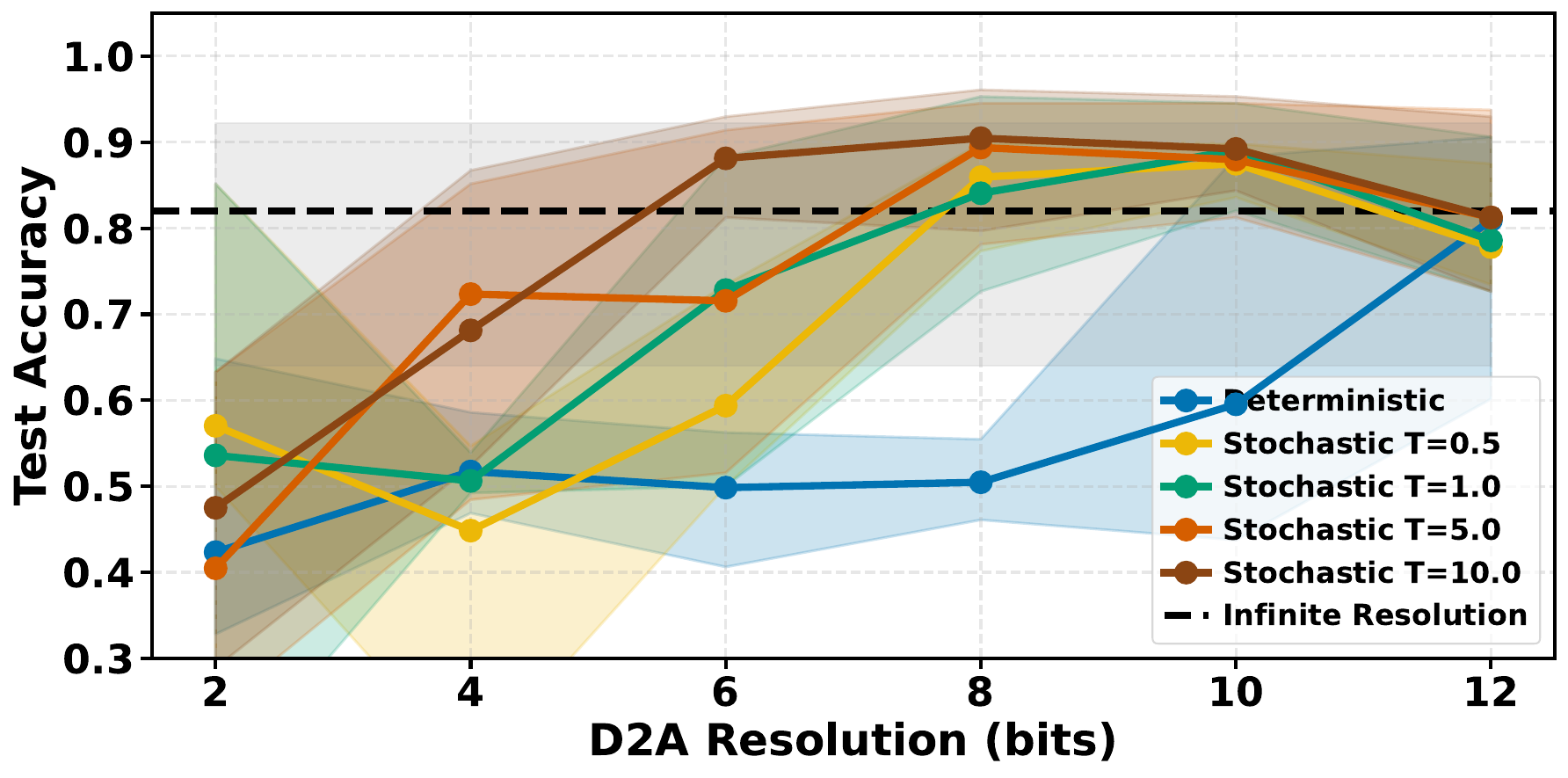}
    \caption{Test accuracy vs Resolution (QNN 2: Breast Cancer).}
    \label{fig:acc_b_bc2}
  \end{subfigure}
  \vspace{-0.1cm}
  \caption{Average train/ test accuracy vs DAC (D2A) resolution on breast cancer dataset (QNN 2), for deterministic and stochastic quantization strategies. Shaded regions show variance across 5 trials.}
  \label{fig:train_test_accuracy_bc2}
\end{figure*}

\section{Results}

This section details our findings across both the experimental paradigms investigated in this work.

\subsection{Inference with Post-Training Quantization}

We first investigate the inference accuracy of pre-trained QNNs (trained with infinite precision) when deployed on quantum computers with finite-resolution control electronics. Figures \ref{fig:ptr_qnn1} and \ref{fig:ptr_qnn2} show the post-training quantization performance of QNN 1 and QNN 2 across all datasets. The figures reveal that test accuracy typically follows a classic elbow curve characteristic, exhibiting improvement with increasing DAC resolution and diminishing returns beyond a dataset-dependent threshold. 

The point of diminishing returns varies by dataset and QNN architecture. For MNIST, both QNNs display this beyond 4 bits, while Fashion-MNIST shows diminishing returns at 3 bits for both QNNs. The Iris dataset exhibits this behavior at 4 bits (QNN 1) and 3 bits (QNN 2). The breast cancer dataset, requires resolutions of 5 bits (QNN 1) and 4 bits (QNN 2) for diminishing returns. Remarkably, even 2-bit DACs recover over 90\% of baseline accuracy (infinite resolution) for MNIST and Fashion-MNIST on QNN 1, though QNN 2 requires 3 bits for comparable recovery.

For practical deployment, 6-bit DACs achieve accuracy indistinguishable from infinite precision baseline across all datasets and both architectures, with the Iris dataset showing saturation earlier at 4 bits (QNN 2) and 5 bits (QNN 1). These results demonstrate that pre-trained QNNs can be reliably deployed on quantum hardware with merely 6-bit control electronics for near-optimal inference accuracy. Furthermore, 4-5 bit DACs suffice to recover over 90\% of baseline performance across all studied configurations, enabling significant power and area reduction in cryo-CMOS control systems with minimal accuracy degradation.

\subsection{QNN Training with Quantization}

When training QNNs directly with finite-resolution DACs using deterministic parameter updates, we observe gradient deadlock at low resolutions. Figure \ref{fig:deadlock_a} reveals that for 2, 4, 6, and 8-bit DACs, the training loss remains constant at a fixed value throughout all epochs. This stagnation occurs because gradient-based parameter updates become smaller than half of the quantization step size ($|\eta \nabla_{\theta}\mathcal{L}| < \Delta/2$), causing parameters to consistently round back to their current quantized values without any effective update (the \textit{gradient deadlock problem}). Even at 10-bit resolution, parameter updates remain marginal and the loss function decays slowly. Only 12-bit DACs enable successful training and although the loss does not fully converge, both training and test accuracies converge and reach values comparable to the infinite-precision baseline (Figure \ref{fig:train_test_accuracy_mnist}).

To overcome gradient deadlock, we introduce controlled stochastic parameter updates that enable training despite sub-(quantization-step) parameter updates. Figure~\ref{fig:deadlock_b} presents training curves for stochastic quantization at temperature $T=1.0$ across all DAC resolutions. At this temperature, 4, 6, 8, and 10-bit systems achieve substantially lower final loss values compared to both 2-bit and 12-bit configurations, indicating that $T=1.0$ is suitable for intermediate resolutions. Unlike conventional smooth loss decay, these training curves exhibit sustained stochasticity throughout the training process, reflecting the probabilistic nature of the parameter update mechanism.

Figures \ref{fig:train_test_accuracy_mnist}, \ref{fig:train_test_accuracy_fmnist}, \ref{fig:train_test_accuracy_iris}, and \ref{fig:train_test_accuracy_bc} compare final training and test accuracies across all DAC resolutions for both deterministic and stochastic quantization strategies at multiple temperatures, showing the performance of QNN 1 on MNIST, Fashion-MNIST, Iris, and breast cancer datasets respectively. Figures \ref{fig:train_test_accuracy_mnist2}, \ref{fig:train_test_accuracy_fmnist2}, \ref{fig:train_test_accuracy_iris2}, and \ref{fig:train_test_accuracy_bc2} present corresponding results for QNN 2.

For deterministic quantization, accuracy exhibits high trial-to-trial variance across all configurations. However, average accuracy shows an increasing trend from 8-12 bits, with improved trainability emerging at 10 bits, indicating that gradient deadlock begins to weaken at this resolution. At 12 bits, deterministic training achieves both mean accuracy and cross-trial variance that matches the infinite-resolution baseline, confirming that 12-bit DACs provide sufficient resolution for conventional gradient-based QNN training without deadlock constraints, which is also supported by Figure \ref{fig:deadlock_a}.

Remarkably, stochastic quantization enables successful training at 4-8 bit resolutions, with performance matching or exceeding infinite-precision baselines at specific temperature configurations. For QNN 1, stochastic parameter updates achieve equal or superior average accuracy with significantly reduced variance across all datasets except breast cancer (Figure \ref{fig:train_test_accuracy_bc})), where 6-10 bit resolutions outperform infinite-resolution training at certain temperatures. Notably, on MNIST and Iris datasets, all temperature configurations surpass infinite-resolution performance at 4-10 bits (Figure \ref{fig:train_test_accuracy_mnist}) and and 6-10 bits (Figure \ref{fig:train_test_accuracy_iris}) respectively, demonstrating that temperature-controlled stochasticity provides exploration noise that enables the optimizer to discover superior regions of the loss landscape compared to gradient descent at infinite precision.

For QNN 2, stochastic quantization exhibits similar performance gains. On MNIST and Iris datasets (Figures \ref{fig:train_test_accuracy_mnist2} and \ref{fig:train_test_accuracy_iris2}), stochastic parameter updates at specific temperatures outperform infinite-resolution training across 4-10 bit resolutions, with minimal cross-trial variability emerging from 6 bits onward. For Fashion-MNIST and breast cancer datasets (Figures \ref{fig:train_test_accuracy_fmnist2} and \ref{fig:train_test_accuracy_bc2}), stochastic training surpasses infinite-resolution performance at 6-10 bits. Notably, lower resolutions benefit from higher temperatures, suggesting that aggressive quantization requires stronger stochasticity for effective parameter updates and exploration of the loss landscape.

However, performance degrades at the resolution extremes. At 2 bits, even stochastic methods yield poor average accuracy, occasionally worse than random guessing and with extreme cross-trial variability across all explored temperature values. At 12 bits, stochastic updates start to underperform suggesting that the fine-grained quantization renders temperature-based exploration counterproductive, introducing unnecessary noise when precise gradient descent is already feasible. This suggests an optimal resolution window (4-10 bits) where temperature-controlled stochasticity maximally benefits training.

This counterintuitive result that QNNs trained with finite resolution DAC constraints can match or exceed infinite-precision performance, demonstrates that constraints of the control electronics need not compromise QML model performance. Instead, appropriately configured quantization can enhance optimization through controlled exploration noise. These findings enable practical hardware-software co-design of QML systems, where 4-10 bit DACs offer substantial power and area savings in cryo-CMOS control electronics while maintaining or improving model performance compared to high-precision alternatives.

\section{Conclusion and Future Work}

This work systematically investigates the interplay between cryogenic control electronics and QML performance through comprehensive evaluation of two QNN architectures across four diverse datasets. We demonstrate that a pre-trained QNN maintains full accuracy when deployed on systems with 6-bit DACs and beyond, with 4-5 bits recovering over 90\% of baseline performance across all configurations, indicating that inference requires minimal control precision. However, training QNNs under quantization constraints reveals gradient deadlock below 12-bit resolution, where parameter updates fall below quantization step sizes. Our temperature-controlled stochastic quantization method resolves this through probabilistic parameter updates, enabling successful training at 4-10 bit resolutions that remarkably matches or exceeds infinite-precision performance which would allow significant reductions in power consumption and silicon area for cryo-CMOS control electronics as quantum computers scale.

Future work should extend these findings across diverse QNN architectures including quantum convolutional neural networks and other QML models such as quantum kernel methods. Results should be further validated on simulators and real quantum devices using parameter-shift gradient estimation, across more diverse dataset benchmarks (including quantum datasets) and machine learning tasks (regression, function approximation, multi-class classification, etc). Furthermore, we plan to systematically optimize temperatures for specific resolutions, and investigate the interplay between DAC quantization and other NISQ-era constraints such as noise and quantum errors.
resolutions.

\bibliographystyle{IEEEtran}
\bibliography{bibliography}

@article{biamonte2017quantum,
  title={Quantum machine learning},
  author={Biamonte, Jacob and others},
  journal={Nature},
  volume={549},
  number={7671},
  pages={195--202},
  year={2017},
  publisher={Nature Publishing Group UK London}
}

@article{schuld2015introduction,
  title={An introduction to quantum machine learning},
  author={Schuld, Maria and others},
  journal={Contemporary Physics},
  volume={56},
  number={2},
  pages={172--185},
  year={2015},
  publisher={Taylor \& Francis}
}

@article{huang2021power,
  title={Power of data in quantum machine learning},
  author={Huang, Hsin-Yuan and others},
  journal={Nature communications},
  volume={12},
  number={1},
  pages={2631},
  year={2021},
  publisher={Nature Publishing Group UK London}
}

@article{huang2022quantum,
  title={Quantum advantage in learning from experiments},
  author={Huang, Hsin-Yuan and others},
  journal={Science},
  volume={376},
  number={6598},
  pages={1182--1186},
  year={2022},
  publisher={American Association for the Advancement of Science}
}

@article{glick2024covariant,
  title={Covariant quantum kernels for data with group structure},
  author={Glick, Jennifer R and others},
  journal={Nature Physics},
  volume={20},
  number={3},
  pages={479--483},
  year={2024},
  publisher={Nature Publishing Group UK London}
}

@article{liu2021rigorous,
  title={A rigorous and robust quantum speed-up in supervised machine learning},
  author={Liu, Yunchao and Arunachalam, Srinivasan and Temme, Kristan},
  journal={Nature Physics},
  volume={17},
  number={9},
  pages={1013--1017},
  year={2021},
  publisher={Nature Publishing Group UK London}
}

@article{huang2025generative,
  title={Generative quantum advantage for classical and quantum problems},
  author={Huang, Hsin-Yuan and others},
  journal={arXiv preprint arXiv:2509.09033},
  year={2025}
}

@article{koczor2024probabilistic,
  title={Probabilistic interpolation of quantum rotation angles},
  author={Koczor, B{\'a}lint and Morton, John JL and Benjamin, Simon C},
  journal={Physical Review Letters},
  volume={132},
  number={13},
  pages={130602},
  year={2024},
  publisher={APS}
}

@article{koczor2024sparse,
  title={Sparse probabilistic synthesis of quantum operations},
  author={Koczor, B{\'a}lint},
  journal={PRX Quantum},
  volume={5},
  number={4},
  pages={040352},
  year={2024},
  publisher={APS}
}

@inproceedings{hu2022quantum,
  title={Quantum neural network compression},
  author={Hu, Zhirui and others},
  booktitle={Proceedings of the 41st IEEE/ACM International Conference on Computer-Aided Design},
  pages={1--9},
  year={2022}
}

@article{PhysRevApplied.12.044054,
  title = {Impact of Classical Control Electronics on Qubit Fidelity},
  author = {van Dijk, J.P.G. and others},
  journal = {Phys. Rev. Appl.},
  volume = {12},
  issue = {4},
  pages = {044054},
  numpages = {20},
  year = {2019},
  month = {Oct},
  publisher = {American Physical Society},
  doi = {10.1103/PhysRevApplied.12.044054},
  url = {https://link.aps.org/doi/10.1103/PhysRevApplied.12.044054}
}

@article{gonzalez2021scaling,
  title={Scaling silicon-based quantum computing using CMOS technology},
  author={Gonzalez-Zalba, MF and others},
  journal={Nature Electronics},
  volume={4},
  number={12},
  pages={872--884},
  year={2021},
  publisher={Nature Publishing Group UK London}
}

@article{sim2019expressibility,
  title={Expressibility and entangling capability of parameterized quantum circuits for hybrid quantum-classical algorithms},
  author={Sim, Sukin and Johnson, Peter D and Aspuru-Guzik, Al{\'a}n},
  journal={Advanced Quantum Technologies},
  volume={2},
  number={12},
  pages={1900070},
  year={2019},
  publisher={Wiley Online Library}
}

@article{bergholm2018pennylane,
  title={Pennylane: Automatic differentiation of hybrid quantum-classical computations},
  author={Bergholm, Ville and others},
  journal={arXiv preprint arXiv:1811.04968},
  year={2018}
}

@article{cerezo2021variational,
  title={Variational quantum algorithms},
  author={Cerezo, Marco and Arrasmith, Andrew and Babbush, Ryan and Benjamin, Simon C and Endo, Suguru and Fujii, Keisuke and McClean, Jarrod R and Mitarai, Kosuke and Yuan, Xiao and Cincio, Lukasz and others},
  journal={Nature Reviews Physics},
  volume={3},
  number={9},
  pages={625--644},
  year={2021},
  publisher={Nature Publishing Group UK London}
}

@article{preskill2018quantum,
  title={Quantum computing in the NISQ era and beyond},
  author={Preskill, John},
  journal={Quantum},
  volume={2},
  pages={79},
  year={2018},
  publisher={Verein zur F{\"o}rderung des Open Access Publizierens in den Quantenwissenschaften}
}

@article{schuld2019evaluating,
  title={Evaluating analytic gradients on quantum hardware},
  author={Schuld, Maria and others},
  journal={Physical Review A},
  volume={99},
  number={3},
  pages={032331},
  year={2019},
  publisher={APS}
}

@article{senokosov2024quantum,
  title={Quantum machine learning for image classification},
  author={Senokosov, Arsenii and others},
  journal={Machine Learning: Science and Technology},
  volume={5},
  number={1},
  pages={015040},
  year={2024},
  publisher={IOP Publishing}
}

@article{chen2024novel,
  title={A novel image classification framework based on variational quantum algorithms},
  author={Chen, Yixiong},
  journal={Quantum Information Processing},
  volume={23},
  number={10},
  pages={362},
  year={2024},
  publisher={Springer}
}

@article{SUN2025130226,
title = {Scalable quantum convolutional neural network for image classification},
journal = {Physica A: Statistical Mechanics and its Applications},
volume = {657},
pages = {130226},
year = {2025},
issn = {0378-4371},
doi = {https://doi.org/10.1016/j.physa.2024.130226},
author = {Yuchen Sun and others}
}

@article{mironowicz2024applications,
  title={Applications of quantum machine learning for quantitative finance},
  author={Mironowicz, Piotr and others},
  journal={arXiv preprint arXiv:2405.10119},
  year={2024}
}

@article{mancilla2022preprocessing,
  title={A preprocessing perspective for quantum machine learning classification advantage in finance using nisq algorithms},
  author={Mancilla, Javier and Pere, Christophe},
  journal={Entropy},
  volume={24},
  number={11},
  pages={1656},
  year={2022},
  publisher={MDPI}
}

@article{doi:10.1021/acs.jcim.1c00166,
author = {Batra, Kushal and others},
title = {Quantum Machine Learning Algorithms for Drug Discovery Applications},
journal = {Journal of Chemical Information and Modeling},
volume = {61},
number = {6},
pages = {2641-2647},
year = {2021},
doi = {10.1021/acs.jcim.1c00166}
}

@article{doi:10.1021/acs.chemrev.4c00678,
author = {Smaldone, Anthony M. and others},
title = {Quantum Machine Learning in Drug Discovery: Applications in Academia and Pharmaceutical Industries},
journal = {Chemical Reviews},
volume = {125},
number = {12},
pages = {5436-5460},
year = {2025},
doi = {10.1021/acs.chemrev.4c00678}
}

@article{deng2012mnist,
  title={The mnist database of handwritten digit images for machine learning research},
  author={Deng, Li},
  journal={IEEE Signal Processing Magazine},
  volume={29},
  number={6},
  pages={141--142},
  year={2012},
  publisher={IEEE}
}

@ARTICLE{8036394,
  author={Patra, Bishnu and others},
  journal={IEEE Journal of Solid-State Circuits}, 
  title={Cryo-CMOS Circuits and Systems for Quantum Computing Applications}, 
  year={2018},
  volume={53},
  number={1},
  pages={309-321},
  doi={10.1109/JSSC.2017.2737549}}

@inproceedings{10.1145/3061639.3072948,
author = {Sebastiano, Fabio and others},
title = {Cryo-CMOS Electronic Control for Scalable Quantum Computing: Invited},
year = {2017},
isbn = {9781450349277},
publisher = {Association for Computing Machinery},
address = {New York, NY, USA},
url = {https://doi.org/10.1145/3061639.3072948},
doi = {10.1145/3061639.3072948},
booktitle = {Proceedings of the 54th Annual Design Automation Conference 2017},
articleno = {13},
numpages = {6},
location = {Austin, TX, USA},
series = {DAC '17}
}

@article{DBLP:journals/corr/abs-1708-07747,
  author    = {Han Xiao and
               Kashif Rasul and
               Roland Vollgraf},
  title     = {Fashion-MNIST: a Novel Image Dataset for Benchmarking Machine Learning
               Algorithms},
  journal   = {CoRR},
  volume    = {abs/1708.07747},
  year      = {2017},
  url       = {http://arxiv.org/abs/1708.07747},
  archivePrefix = {arXiv},
  eprint    = {1708.07747},
  bibsource = {dblp computer science bibliography, https://dblp.org}
}

@article{Fisher1936THEUO,
  title={THE USE OF MULTIPLE MEASUREMENTS IN TAXONOMIC PROBLEMS},
  author={Rory A. Fisher},
  journal={Annals of Human Genetics},
  year={1936},
  volume={7},
  pages={179-188},
  url={https://api.semanticscholar.org/CorpusID:29084021}
}

@inproceedings{Street1993NuclearFE,
  title={Nuclear feature extraction for breast tumor diagnosis},
  author={William Nick Street and William H. Wolberg and Olvi L. Mangasarian},
  booktitle={Electronic imaging},
  year={1993},
  url={https://api.semanticscholar.org/CorpusID:14922543}
}

@article{havlivcek2019supervised,
  title={Supervised learning with quantum-enhanced feature spaces},
  author={Havl{\'\i}{\v{c}}ek, Vojt{\v{e}}ch and others},
  journal={Nature},
  volume={567},
  number={7747},
  pages={209--212},
  year={2019},
  publisher={Nature Publishing Group UK London}
}

@article{perez2020data,
  title={Data re-uploading for a universal quantum classifier},
  author={P{\'e}rez-Salinas, Adri{\'a}n and others},
  journal={Quantum},
  volume={4},
  pages={226},
  year={2020},
  publisher={Verein zur F{\"o}rderung des Open Access Publizierens in den Quantenwissenschaften}
}

@article{bowles2024better,
  title={Better than classical? the subtle art of benchmarking quantum machine learning models},
  author={Bowles, Joseph and Ahmed, Shahnawaz and Schuld, Maria},
  journal={arXiv preprint arXiv:2403.07059},
  year={2024}
}

@article{phalak2023shot,
  title={Shot optimization in quantum machine learning architectures to accelerate training},
  author={Phalak, Koustubh and Ghosh, Swaroop},
  journal={IEEE Access},
  volume={11},
  pages={41514--41523},
  year={2023},
  publisher={IEEE}
}

@article{acedo2025pulsed,
  title={Pulsed learning for quantum data re-uploading models},
  author={Acedo, Ignacio B and others},
  journal={arXiv preprint arXiv:2512.10670},
  year={2025}
}

@article{zhuang2024non,
  title={Non-hemolytic peptide classification using a quantum support vector machine},
  author={Zhuang, Shengxin and others},
  journal={arXiv preprint arXiv:2402.03847},
  year={2024}
}

@ARTICLE{10287646,
  author={Pérez-Bailón, Jorge and others},
  journal={IEEE Transactions on Instrumentation and Measurement}, 
  title={Cryogenic Measurement of CMOS Devices for Quantum Technologies}, 
  year={2023},
  volume={72},
  number={},
  pages={1-7},
  doi={10.1109/TIM.2023.3325446}}

@ARTICLE{701179,
  author={Treadgold, N.K. and Gedeon, T.D.},
  journal={IEEE Transactions on Neural Networks}, 
  title={Simulated annealing and weight decay in adaptive learning: the SARPROP algorithm}, 
  year={1998},
  volume={9},
  number={4},
  pages={662-668},
  doi={10.1109/72.701179}}

@book{Goodfellow-et-al-2016,
    title={Deep Learning},
    author={Ian Goodfellow and Yoshua Bengio and Aaron Courville},
    publisher={MIT Press},
    note={\url{http://www.deeplearningbook.org}},
    year={2016}
}

@article{10.5555/329748.329752,
author = {Sexton, Randall S. and Dorsey, Robert E. and Johnson, John D.},
title = {Beyond back propagation: using simulated annealing for training neural networks},
year = {1999},
issue_date = {July/Sept. 1999},
publisher = {IGI Global},
address = {USA},
volume = {11},
number = {3},
issn = {1063-2239},
journal = {J. End User Comput.},
month = jul,
pages = {3–10},
numpages = {8}
}

@Article{Rumelhart1986,
author={Rumelhart, David E.
and Hinton, Geoffrey E.
and Williams, Ronald J.},
title={Learning representations by back-propagating errors},
journal={Nature},
year={1986},
month={Oct},
day={01},
volume={323},
number={6088},
pages={533-536},
issn={1476-4687},
doi={10.1038/323533a0}
}

\end{document}